\documentclass[lettersize,journal]{IEEEtran}
\usepackage{amsmath,amsfonts}
\usepackage{algorithmic}
\usepackage{algorithm}
\usepackage{array}
\usepackage[caption=false,font=normalsize,labelfont=sf,textfont=sf]{subfig}
\usepackage{textcomp}
\usepackage{stfloats}
\usepackage{url}
\usepackage{verbatim}
\usepackage{graphicx}
\usepackage{cite}
\usepackage[switch]{lineno} 
\usepackage{color}
\usepackage{fancyhdr}
\begin{document}

\title{\textcolor{black}{Intensity-Adjustable Non-contact Cold Sensation Presentation Based on the Vortex Effect}}

\author{Jiayi~Xu,~\IEEEmembership{Student Member,~IEEE},  Shunsuke~Yoshimoto,~\IEEEmembership{Member,~IEEE}, Naoto~Ienaga,~\IEEEmembership{Member,~IEEE} and~Yoshihiro~Kuroda,~\IEEEmembership{Member,~IEEE}
\thanks{Manuscript received April xx, xxxx; revised August xx, xxxx. This work was supported in part by grants from JSPS KAKENHI (JP21H03474, JP21K19778) and in part by JST SPRING (JPMJSP2124).}
\thanks{J. Xu is with the Graduate School of Science and Technology, Degree Programs in Systems and Information Engineering, University of Tsukuba, Japan (e-mail: xujiayi@le.iit.tsukuba.ac.jp).}
\thanks{S. Yoshimoto is with the Department of Human \& Engineered Environmental Studies, The University of Tokyo, Japan (e-mail: yoshimoto@k.u-tokyo.ac.jp).}
\thanks{N. Ienaga and Y. Kuroda are with the Faculty of Engineering, Information and Systems, University of Tsukuba, Japan (e-mail: ienaga@iit.tsukuba.ac.jp; kuroda@iit.tsukuba.ac.jp).}}


\maketitle
\thispagestyle{fancy}
\begin{abstract}
\textcolor{black}{Cold sensations of varying intensities are perceived when human skin is subject to diverse environments. The accurate} presentation of temperature changes is important to elicit immersive sensations in applications such as virtual reality. We developed a method to elicit intensity-adjustable non-contact cold sensations based on the vortex effect. We applied \textcolor{black}{this effect} to generate cold air at approximately 0~$^\circ$C and varied the skin temperature over a wide range. The perception of different temperatures can be elicited by adjusting \textcolor{black}{the volume flow rate of the cold air}. Additionally, we introduced a cooling model to relate the changes in skin temperature to various parameters such as the cold air volume flow rate and distance from the cold air outlet to the skin. For validation, we conducted measurement experiments and found that our model can estimate the change in skin temperature with a root mean-square error of 0.16~$^\circ$C. Furthermore, we evaluated the performance of a prototype in psychophysical cold discrimination experiments based on the discrimination threshold. Thus, cold sensations of varying intensities can be generated by varying the parameters. These cold sensations can be combined with images, sounds, and other stimuli to create an immersive and realistic artificial environment.
\end{abstract}


\section{Introduction}
\IEEEPARstart{I}{n} human thermal sensing, cold is of particular importance because humans are more sensitive to decreases in skin temperature than increases~\cite{Somatosensory1998}. The ability to perceive cold enables us to interact with our surroundings~\cite{warmAndCool}. Meanwhile, with the advancement of virtual reality (VR) technology and the emergence of the Metaverse, new methods of recreating interactions are expected. By presenting cold sensations with adjustable intensity, we can create a new way of interaction. For example, various artificial environments, such as a coast with a breeze or \textcolor{black}{a} snow field with a strong wind, can be created to provide users with immersive experiences. Similarly, some of the real-life sensations, such as the cooling sensation when entering an air-conditioned room or the blowing of cold air when the refrigerator is opened, can be recreated. As these sensations are not accompanied by a sense of touch, a method that can present intensity-adjustable non-contact cold sensations is essential.
\begin{figure}[t]
	\centering
	\subfloat[]{\includegraphics[scale=0.3]{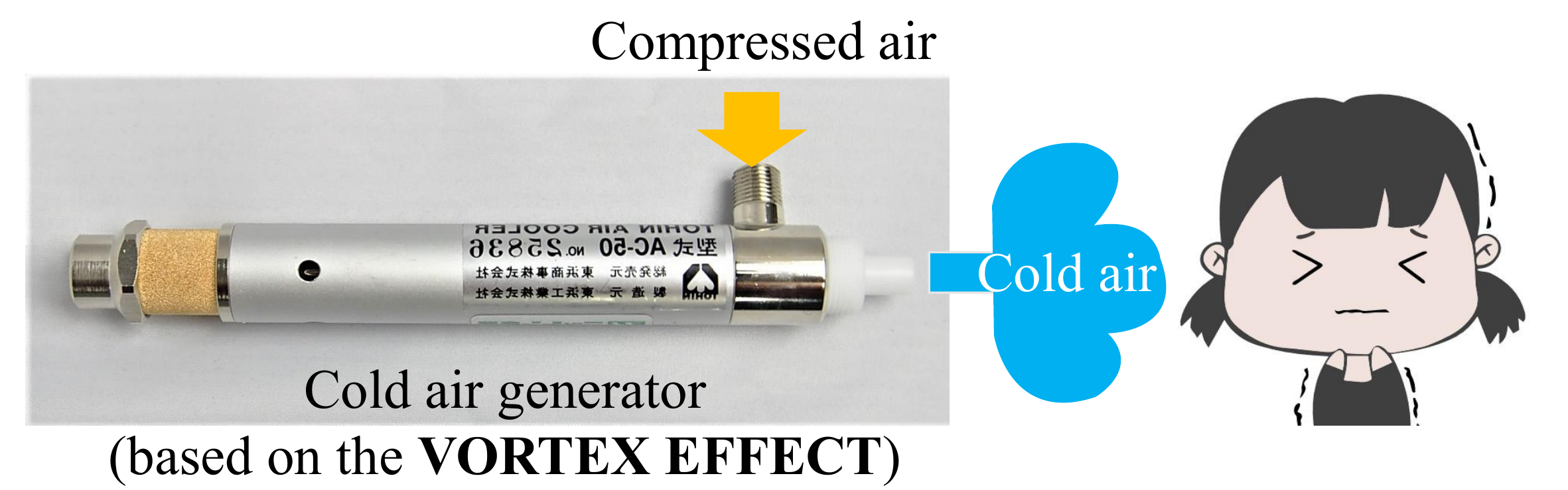}}
    \hfil
    \subfloat[]{\includegraphics[scale=0.3]{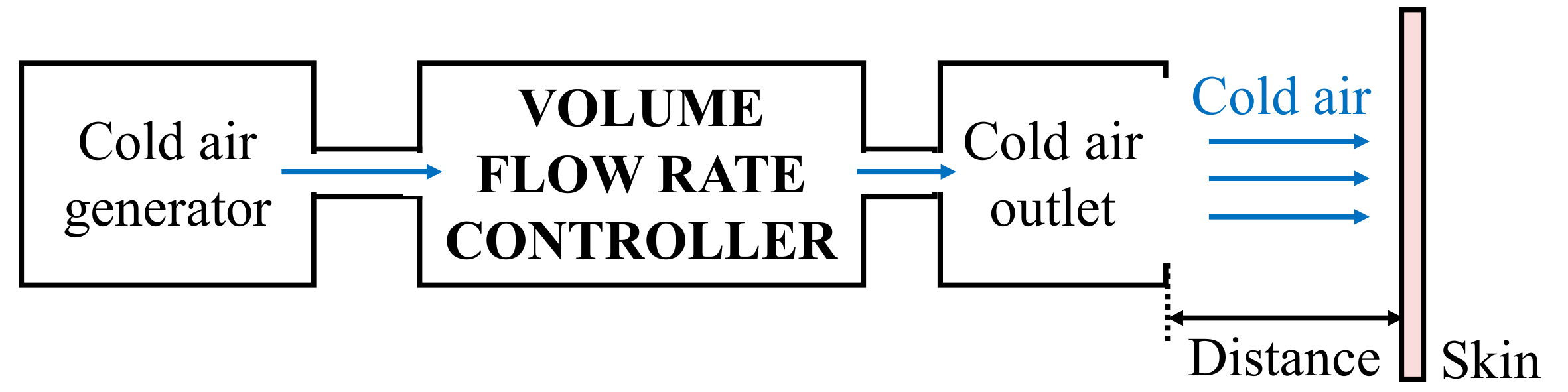}}
    \hfil
    \subfloat[]{\includegraphics[scale=0.3]{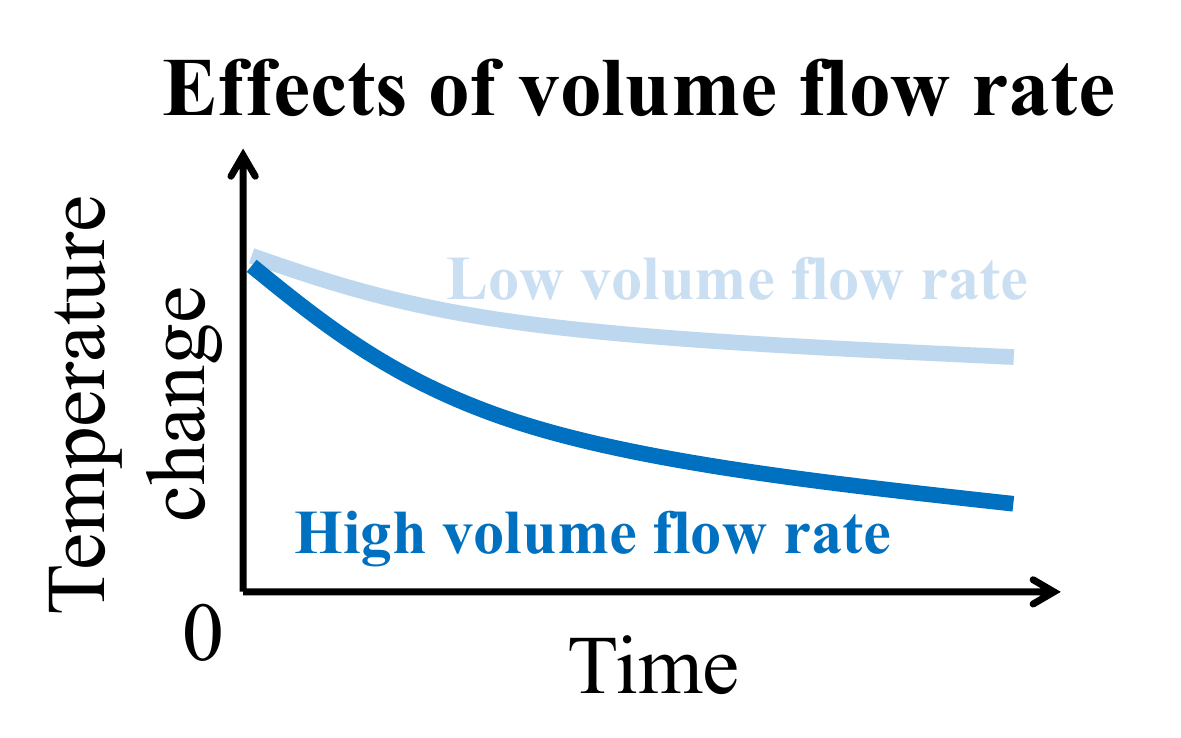}}
    \hfil
    \subfloat[]{\includegraphics[scale=0.3]{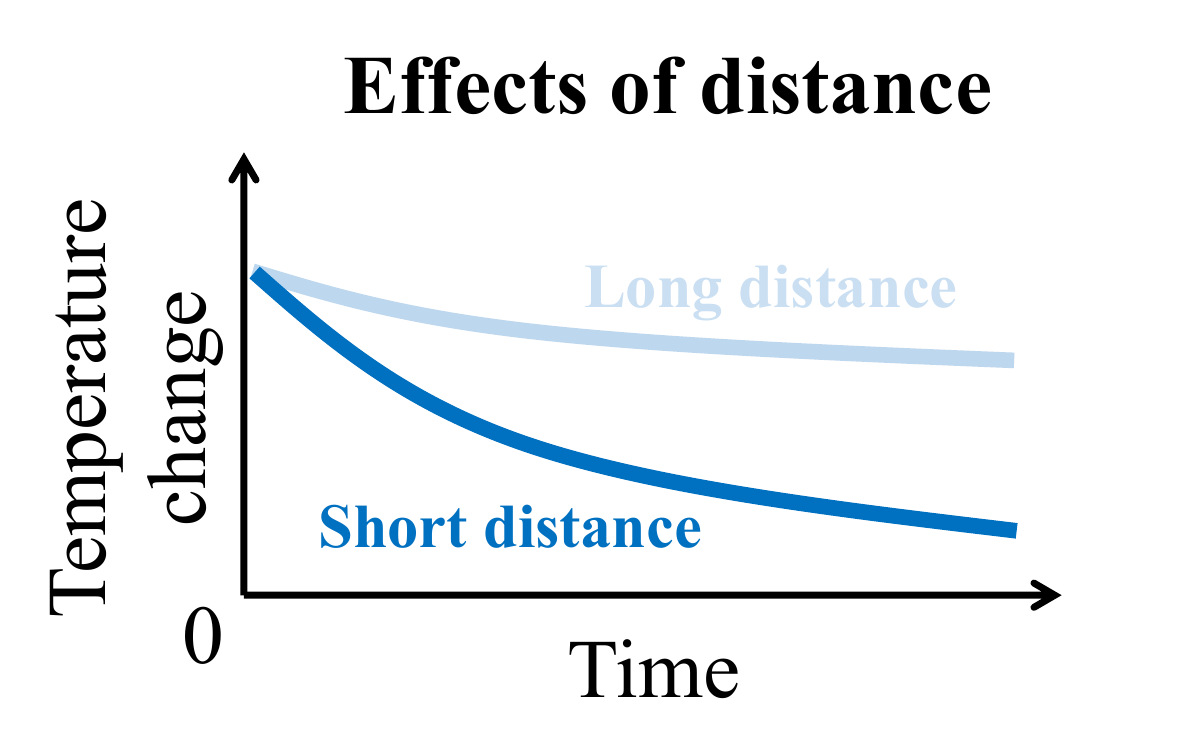}}
	\caption{Overview of proposed method. We leverage the vortex effect to generate low-temperature cold air (a). We control the intensity of cold sensation and temperature perception by varying the volume flow rate of cold air (b). As both the volume of generated cold air (c) and the distance between the cold air outlet and skin (d) influence temperature, we use a model relating these factors with skin temperature for proper adjustment.}
	\label{fig:Overview}
\end{figure}
\par Cold sensations are closely related to changes in skin temperature~\cite{temperatureChangeRate,thresholds}, and a greater change can elicit a stronger cold sensation, whereas a smaller change \textcolor{black}{results in a less intense cold sensation}. Furthermore, according to the cold perception threshold, when the rate of temperature change is approximately 0.3~$^\circ$C/s, humans are able to perceive cold within approximately one second~\cite{thresholds}. Therefore, first, for presenting different intensities of cold sensations, a cooling method capable of causing a temperature change rate not less than 0.3~$^\circ$C/s should be developed. Second, unlike the contact-type method, which alters the skin temperature by controlling the temperature of the contact surface (medium), \textcolor{black}{a} non-contact cooling method cannot control skin temperature directly. Therefore, a model should be determined to relate skin temperature changes to various influencing factors, aiming to characterize and control these relations.
\par For non-contact cooling, Nakajima \textit{et al.}~\cite{mist0,mist1,mist2} proposed an effective method that uses ultrasound to transport mist to the skin. The focused ultrasound would accelerate the evaporation of mist on the skin to create a cooling spot, achieving a maximum temperature change of 4.6~$^\circ$C in 1~s. \textcolor{black}{The dominant factor enabling this method to affect temperature change is evaporation.} In some other cooling methods, convection is the main factor affecting the temperature change, such as air conditioning. However, it requires time to generate sufficient cold air to reduce the skin temperature. Alternatively, fans, blowers, or cold air generated with dry ice can be used to elicit a non-contact cold sensation~\cite{fan,fan1,dryIce}. However, these methods have various shortcomings, such as the inability to notably reduce the skin temperature due to the long time required to precool air. 
\par \textcolor{black}{This study aims to develop an intensity-adjustable non-contact cooling method, which relies on convection as the dominant factor affecting the temperature change.} To this end, we propose an intensity-adjustable thermal display based on the vortex effect, which is the thermal separation of swirling air in a tube. The vortex effect was discovered in thermal engineering and is generated by compressed air injected into a tube, causing it to swirl and separate into cold and hot air~\cite{vortextube}. Unlike traditional air conditioning, the vortex effect can generate low-temperature cold air immediately. Therefore, we consider \textcolor{black}{this technique} to be suitable for the proposed method. As shown in Fig.~\ref{fig:Overview} (a), we transfer the generated cold air to the skin to vary its temperature. We then adjust the intensity of cold sensations by changing the volume flow rate of the cold air to elicit different temperature changes on the skin, as shown in Fig.~\ref{fig:Overview} (b). \textcolor{black}{Our presentation component, i.e., the cold air outlet (a round tube with a diameter of 8~mm), is lightweight} and can be miniaturized and customized for easy incorporation into VR wearable devices. In addition, the compressor can be placed in a remote location to \textcolor{black}{avoid} problems related to size and noise. In addition, we introduce a cooling model that describes the proposed method and relates skin temperature changes with their influencing factors. The model allows us to design temperature changes and elicit those changes as expected (see Section \uppercase\expandafter{\romannumeral3}).
\par We thoroughly evaluated the proposed method for non-contact intensity-adjustable cold sensation presentation to guide the development of thermal displays. Based on a preliminary investigation, we found that skin temperature changes and cold sensations are related to the volume flow rate of cold air generated using the vortex effect~\cite{coldAir}. Furthermore, the distance between the cold air outlet and skin may influence the perception of cold and temperature changes. To analyze these factors, we developed a cooling model considering the influence of the display distance on skin temperature. Additionally, we evaluated temperature changes and cold sensations according to the volume flow rates at different distances. Moreover, we conducted experiments to evaluate a prototype and the proposed model (see Section \uppercase\expandafter{\romannumeral4}) and investigated the responses of the proposed method on human perception (see Section \uppercase\expandafter{\romannumeral5}). The experimental results allowed us to determine the temperature changes and cold sensations that our system can provide as well as thresholds for eliciting non-contact cold sensations. 

\section{Related Works}
\par In this section, we first introduce the perceptual characteristics of cold sensations. We then review related works on eliciting cold sensations. Finally, we detail the contributions of our study.
\subsection{Perceptual Characteristics of Cold Sensations}
\par The mechanism in humans for sensing temperature has remained a mystery for a long time. With the discovery of transient receptor potential (TRP) channels in 1997, remarkable progress has been made toward unveiling this mechanism~\cite{thermoreceptors0}. There are 27 TRP channels in humans, of which nine are sensitive to temperature~\cite{thermoreceptors1,thermoreceptors2}. Upon activation, TRP channels increase the excitability of thermoreceptors, through \textcolor{black}{which humans perceive} cold.
\par \textcolor{black}{Thermoreceptors sensitive to cold can encode a wide range of temperatures and are maximally responsive between 22 to 28 $^\circ$C for cold perception~\cite{thermoreceptors,temperatureReceptors}. Thermoreceptors are also sensitive to changes in skin temperature and are affected by the changing rate and range.} Therefore, temperature sensitivity is closely related to temperature changes in the skin~\cite{temperatureChangeRate,thresholds}. If the temperature changes more rapidly, the cold sensation becomes stronger.
\subsection{Eliciting Cold Sensations}
\par Since the 1990s, Peltier elements have been used to evoke cold sensations~\cite{peltier0,peltier1,peltier2,peltier3,peltier4,peltier5,peltier6,peltier7,peltier8,peltier9}. Yamamoto \textit{et al.}~\cite{peltier2} used Peltier elements to simulate temperature changes when touching objects to elicit the perception of different materials. To accurately adjust the sensation intensity, they developed a model for predicting temperature changes. Peiris \textit{et al.}~\cite{peltier6} developed a head-mounted display that elicits hot and cold sensations in response to the direction of the user’s head. This is achieved by placing multiple Peltier elements around the headpiece. Kushiyama \textit{et al.}~\cite{peltier9} combined images, music, and hot and cold sensations to realize a new form of artistic expression. 
\par Various studies have used water to elicit cold sensations. Sakaguchi \textit{et al.}~\cite{water1,water2} used a valve to alternate hot and cold water in a thermal display to switch the presented sensation. G{\"u}nther \textit{et al.}~\cite{water3} adjusted the intensity of cold sensations by changing the temperature of water. On the other hand, Cai \textit{et al.}~\cite{air} used air instead of water as the heat source. \textcolor{black}{The intensity of cold sensations can be controlled by adjusting the temperature of air injected into a thermal display placed in contact with the skin. However, all these methods require the skin to be in direct contact with the thermal display.}
\par Although contact thermal displays reproduce touch sensations, creating various artificial environments, \textcolor{black}{such as a coast with a breeze or a snow field} with a strong wind, requires non-contact cold presentations. \textcolor{black}{To generate these, Nakajima \textit{et al.}~\cite{mist0,mist1,mist2} developed a cooling method with evaporation as the main factor of temperature change. By using focused ultrasound, they transported a mist to the skin to cool it. Other} studies have focused on cooling methods with convection as the main factor. To generate cold sensations with this mechanism, Dionisio \textit{et al.}~\cite{fan} used a fan as a first attempt. However, Ogasahara \textit{et al.}~\cite{fan1} found that a fan achieved a maximum temperature change rate of only 0.05~$^\circ$C/s, which only elicits a weak cold sensation. Nakajima \textit{et al.}~\cite{dryIce} remotely elicited cold sensations using a focused ultrasound to transfer cooled air generated by dry ice to the skin, \textcolor{black}{but this approach required a period of time to generate precooled air.}

\subsection{Contributions}
\par The contributions of this study are as follows:
\begin{itemize}
    \item \textcolor{black}{We developed a non-contact intensity-adjustable cooling method based on the vortex effect with convection as the dominant factor to elicit cold sensations.}
    \item As the vortex effect \textcolor{black}{can generate} low-temperature cold air immediately from compressed air, the generated cold air can be contactlessly transferred to quickly alter the skin temperature and elicit a strong cold sensation. Moreover, changing the volume flow rate of cold air allows the sensation intensity to be easily adjusted. 
    \item \textcolor{black}{Our presentation component, i.e., a cold air outlet (a round tube with a diameter of 8~mm), is lightweight and can be miniaturized and customized for easy incorporation into VR wearable devices. In addition, the compressor can be placed in a remote location to avoid problems related to size and noise.}
\end{itemize}

\section{Method to Elicit Cold Sensations}
\par In this section, we first explain the vortex effect. We then introduce the proposed intensity-adjustable non-contact method for eliciting cold sensations based on the vortex effect and describe its design and implementation. Finally, we detail the proposed cooling model that relates the change in skin temperature to its influencing factors.
\subsection{Vortex Effect}
\par With the vortex effect, heat in a spiral flow through a tube is separated into hot and cold airflows~\cite{vortextube}. This effect can be used as a low-temperature heat source because it provides low-temperature cold air instantly from a compressed air supply. Thus, we incorporate the vortex effect into a thermal display to achieve stable and rapid cooling.
\par
\begin{figure}[t]
	\centering
	\includegraphics[scale=0.3]{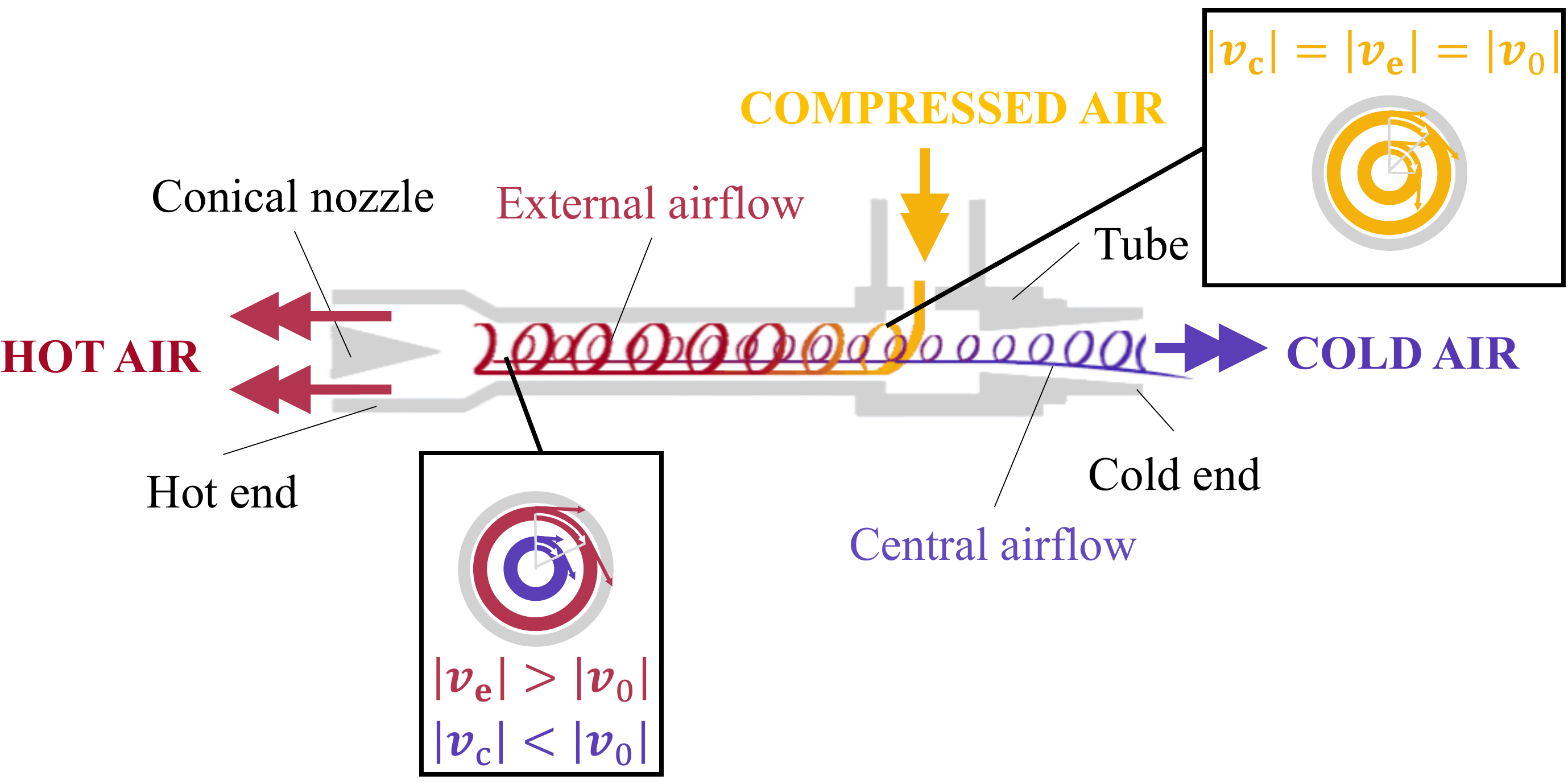}
	\caption{Heat separation due to vortex effect.}
	\label{fig:VortexEffect}
\end{figure}

 Fig.~\ref{fig:VortexEffect} illustrates heat separation caused by the vortex effect. When compressed air is injected into a tube, it becomes a high-speed rotating airflow comprising external and central airflows with the same velocity, $v_{0}$. Then, the airflow moves toward the hot end. In this process, the external airflow forces the central airflow to rotate at a constant angle, reducing \textcolor{black}{the} velocity $v_{\rm c}$ of the central airflow and increasing \textcolor{black}{the} velocity $v_{\rm e}$ of the external airflow. The kinetic energy \textcolor{black}{is converted to} heat and propagates outwards from the central airflow. As the hot end has a conical nozzle, only the external airflow (hot air) exits through this end. The central airflow (cold air) is pushed back into the center of the tube by the reaction force of the conical nozzle and exits through the cold end. 
\par The temperature of the generated cold air is determined by the pressure and temperature of the supplied air. The temperature of the cold air decreases as the air pressure increases or the temperature of the supplied air decreases. Moreover, colder air can be output by varying the position of the conical nozzle at the hot end to increase the amount of hot air, but the volume of cold air becomes small. Furthermore, as more hot air is discharged, \textcolor{black}{a greater amount of heat is dissipated.}
\subsection{System Design and Implementation}
\par We propose a novel intensity-adjustable non-contact method for cold sensation presentation based on the vortex effect. The vortex effect generates low-temperature cold air and allows the adjustment of the intensity of the cold sensation by varying the volume flow rate of cold air without changing its temperature. The fixed temperature overcomes limitations related to changing the air temperature. According to the characteristics of the vortex effect, the temperature of the generated cold air can be altered according to the pressure or temperature of the supplied air. However, it takes time to alter these parameters. On the other hand, moving the conical nozzle at the hot end can alter the temperature of the generated cold air. However, repeated motion operations can destroy the tube structure. 
\par
\begin{figure}[t]
	\centering
	\subfloat[]{\includegraphics[scale=0.26]{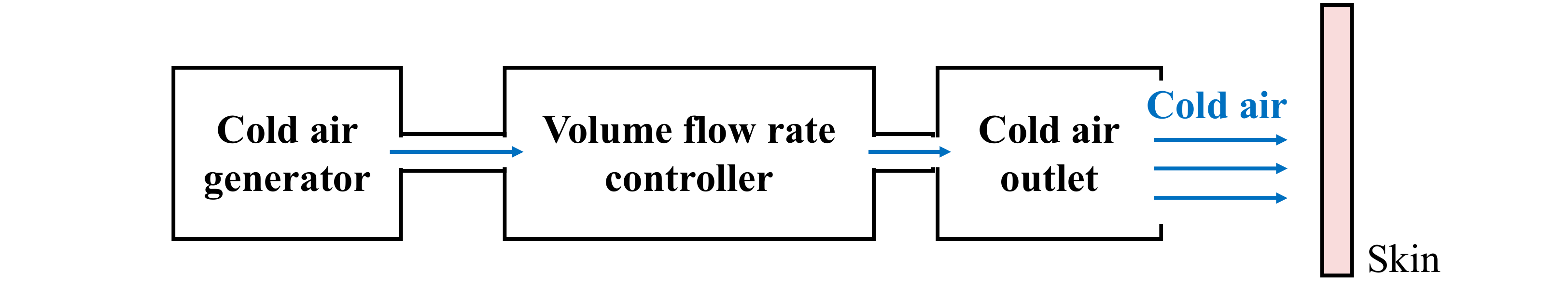}}
	\hfil
	\subfloat[]{\includegraphics[scale=0.26]{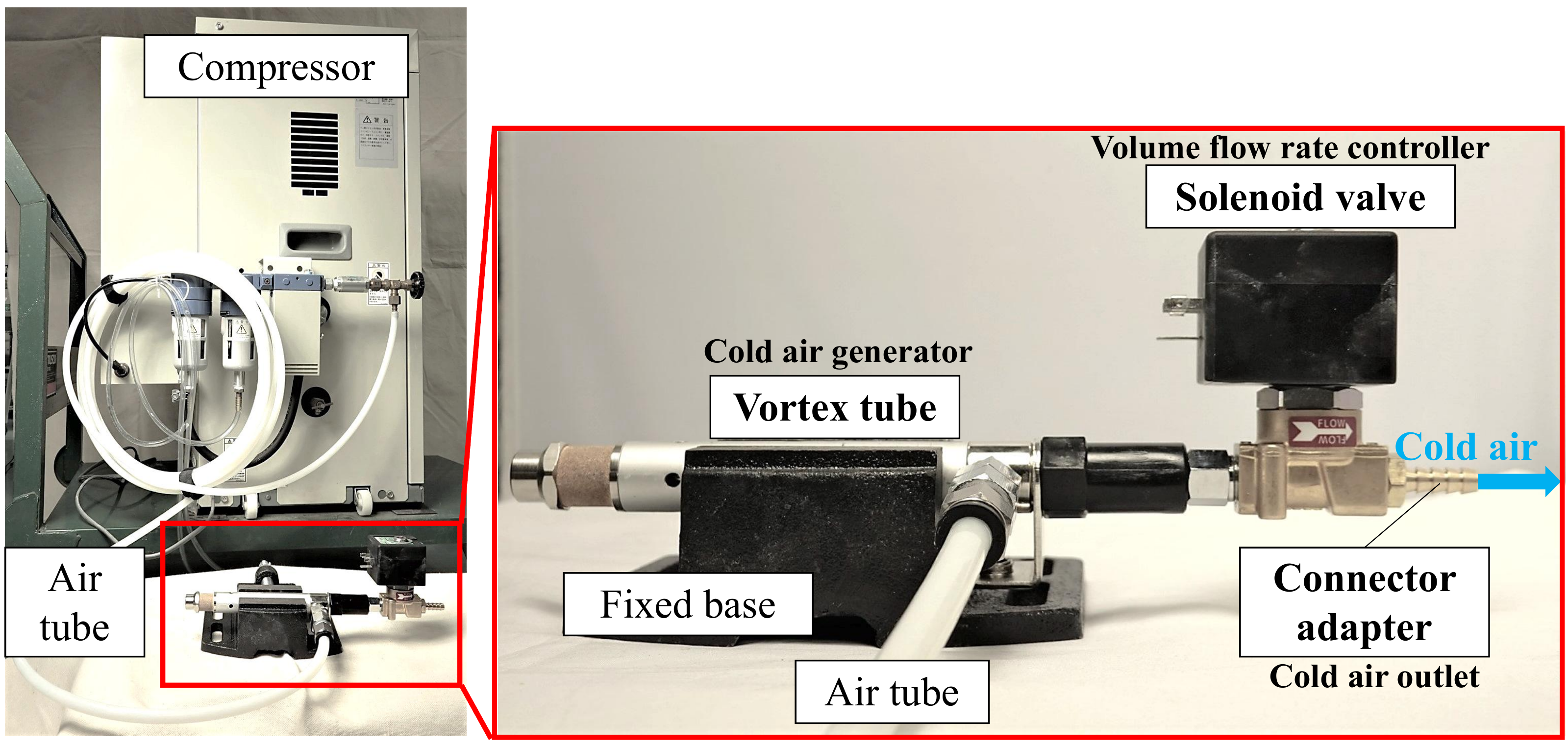}}	
	\caption{Intensity-adjustable non-contact cold sensation presentation. Schematic of proposed system (a) and implemented prototype (b).}
	\label{fig:PrototypeSystem}
\end{figure}

 Fig.~\ref{fig:PrototypeSystem} (a) shows the schematic of the developed prototype. First, a cold air generator produces cold air at a constant volume flow rate. \textcolor{black}{Next, the cold air is regulated by adjusting the applied voltage to the volume flow rate controller.} Finally, the cold air outlet of the system delivers regulated cold air. Fig.~\ref{fig:PrototypeSystem} (b) shows the developed system comprising a cold air generator, a volume flow rate controller, and a cold air outlet. These elements are mounted using a vortex tube (Tohin AC-50), a solenoid valve (ASCO Positive-flow-202), and a connector adapter (UXCELL B07GYRBYMW), respectively. Additionally, an air compressor (Hitachi POD-0.75LES) provides air to the tube. As the air compressor in the system can output air at pressures of 0.6 to 0.8~MPa, the temperature of the cold air at the tube outlet is constant.
\par
\begin{figure}[t]
	\centering
	\includegraphics[scale=0.28]{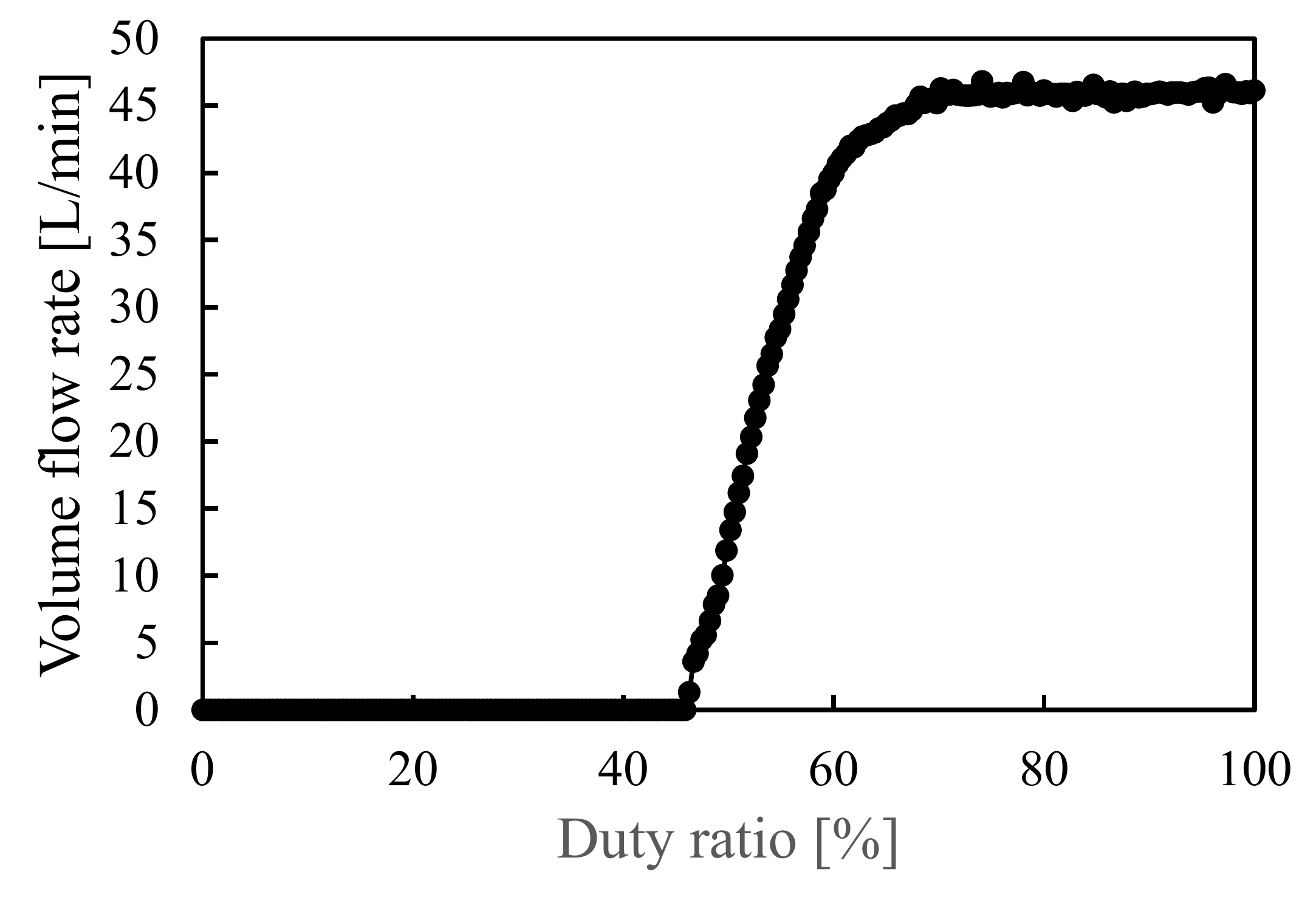}
	\caption{Output volume flow rate according to pulse width modulation duty ratio. }
	\label{fig:FlowVolumeWithDutyRatio}
\end{figure}

 By using pulse width modulation (PWM), we control the solenoid valve via serial communications using a microcomputer (Arduino Due). A high-power metal–oxide–semiconductor field-effect transistor trigger switch drive module (Baoblaze B07BVKL4VW) performs PWM control. Fig.~\ref{fig:FlowVolumeWithDutyRatio} shows the \textcolor{black}{relationship} between the PWM duty ratio and output volume flow rate, which is determined using an amplifier separation gas flow sensor (Keyence FD-A50). The output volume flow rate ranges from 0 to 45~L/min.
\subsection{Cooling Model}
\par
\begin{figure}[t]
    \centering
        \subfloat[]{\includegraphics[width=0.96\hsize]{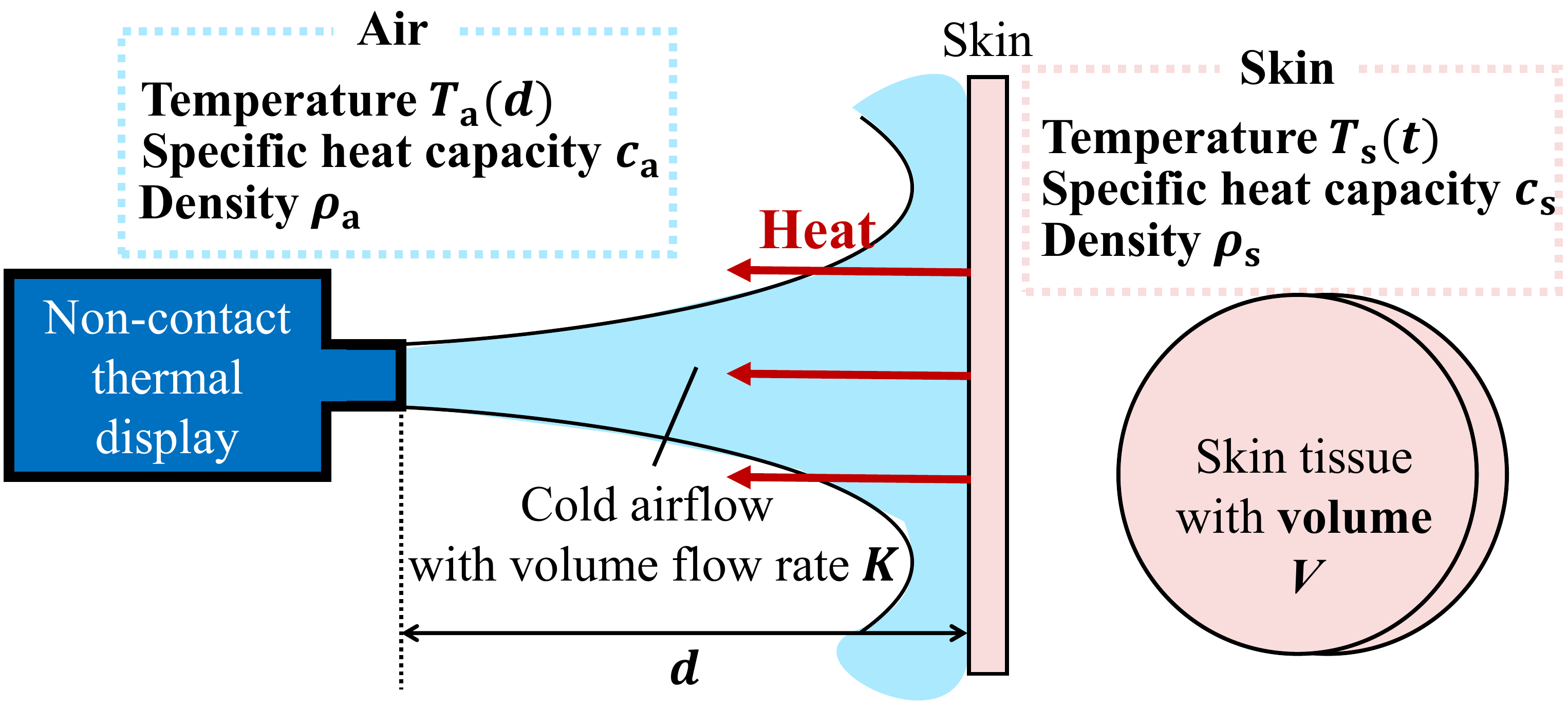}}
        \hfil
        \subfloat[]{\includegraphics[width=0.35\hsize]{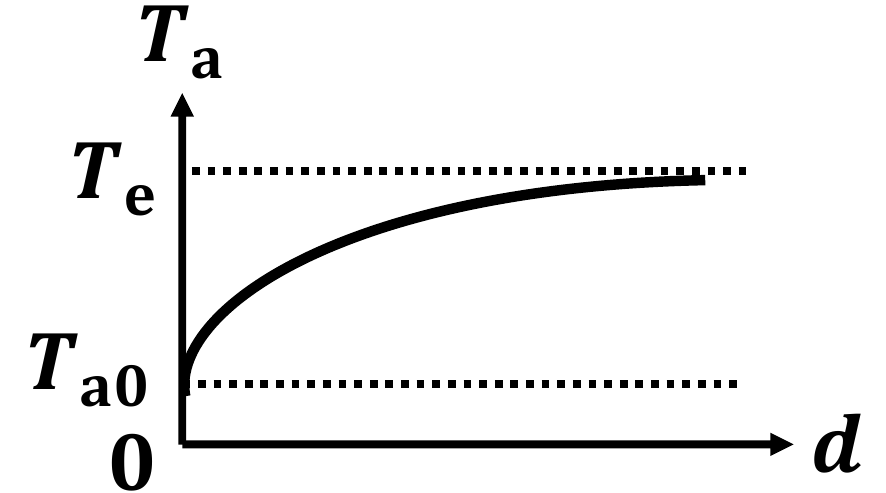}} 
        \subfloat[]{\includegraphics[width=0.35\hsize]{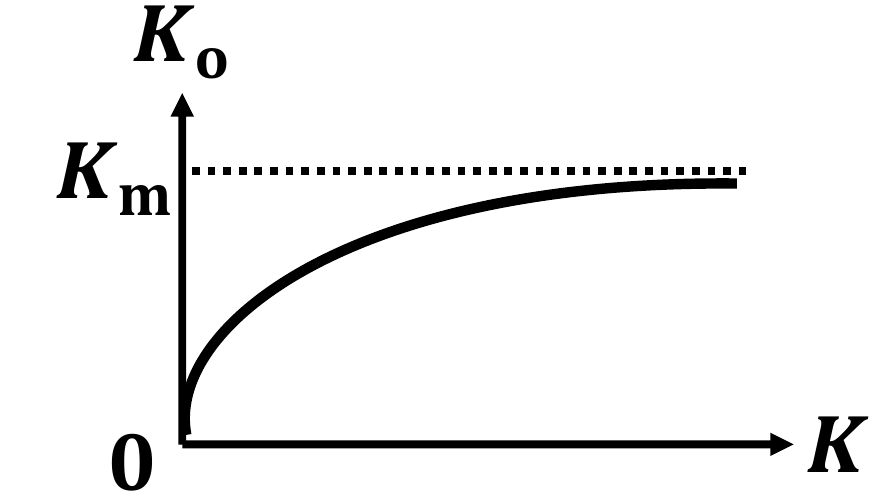}}  \\ 
        \caption{Cooling model with diagram of cooling process (a). Temperature of cold air according to distance $d$ from cold air outlet to skin (b). Volume flow rate of cold air in contact with skin (c).}
    \label{fig:CoolingModel}
\end{figure}

 We propose a cooling model that relates the change in skin temperature to both the volume flow rate of cold air and the distance between the cold air outlet and the skin. Fig.~\ref{fig:CoolingModel}(a) illustrates skin cooling using the proposed system. \textcolor{black}{When the cold air makes contact with the skin, it absorbs heat from the skin, and changes the skin temperature.} Additionally, the exchange of heat between the blood and skin and the production of heat by the body are factors that affect skin temperature~\cite{Pennes}. For a concise explanation of how cold air affects changes in temperature, we exclude the bodily factors and introduce a cooling model based on the following assumptions: 
\begin{enumerate}
    \item As the specific heat capacity of air is smaller than that of the skin, cold air quickly transfers its heat to the skin. When cold air contacts the skin, the air temperature quickly increases to become equal to that of the skin. In addition, the skin only absorbs part of the heat provided by the cold air with an absorption ratio of $\sigma(0~\leq~\sigma~\leq~1)$.
    \item The temperature of the cold air reaching the skin increases with distance $d$ from the cold air outlet. As shown in Fig.~\ref{fig:CoolingModel}(b), when the distance is zero, cold air shows an initial temperature of $T_{\rm a0}$, and when the distance is sufficiently large, cold air heats up to an ambient temperature of $T_{\rm e}$.
    \item There is a maximum volume flow rate, $K_{\rm m}$, of cold air that can contact the skin, as shown in Fig.~\ref{fig:CoolingModel}(c). When the output volume flow rate, $K$, of cold air is high, cold air reaching the skin first contacts the skin. Thus, the cold air that follows cannot contact the skin and dissipates into the surrounding ambient air. 
    \item The volume flow rate and distance alter the skin area in direct contact with the cold air, but heat diffusion hinders the determination of the contact area. In addition, cold sensation does not solely depend on the change in temperature at a particular point on the skin; \textcolor{black}{it depends on the area of skin cooled~\cite{area1}}. Therefore, we consider the average temperature change in a skin area of fixed volume, $V$. In subsequent experiments, we set a filter to maintain the constant size of the area.
    Regarding the thermal properties of the skin, we assume that the tissue is isotropic and uniform. Accordingly, we consider \textcolor{black}{the} density $\rho_{\rm s}$ and specific heat capacity $c_{\rm s}$ \textcolor{black}{to be} constant.
\end{enumerate}
Based on assumption 2, temperature $T_{\rm a}(d)$ of cold air reaching the skin is given by
\begin{equation}\label{eq:Tad}
    T_{\rm a}(d)=T_{\rm a0}+(T_{\rm e}-T_{\rm a0})(1-e^{-\alpha d}),
\end{equation}
where $\alpha(\alpha>0)$ is a constant. Based on assumption 3, volume flow rate $K_{\rm o}(K)$ of cold air in contact with the skin is given by
\begin{equation}\label{eq:Ko}
    K_{\rm o}(K)=K_{\rm m}(1-e^{-\theta K}),
\end{equation}
where $\theta(\theta>0)$ is a constant. Based on assumption 1, heat $Q_{\rm c}$ provided by cold air per unit time is given by
\begin{equation}\label{eq:Qc}
    Q_{\rm c}=-\rho_{\rm a}c_{\rm a}K_{\rm o}(K)(T_{\rm s}(t)-T_{\rm a}(d)),
\end{equation}
where $c_{\rm a}$ and $\rho_{\rm a}$ are the specific heat capacity and density of air, respectively. The actual heat absorbed by the skin per unit time, $Q_{\rm s}$, is given by
\begin{equation}\label{eq:Qs}
    Q_{\rm s}=\sigma Q_{\rm c}.
\end{equation}
The heat change per unit time has the following relationship with the skin temperature change:
\begin{equation}\label{eq:Qs1}
    Q_{\rm s}=c_{\rm s}\rho_{\rm s}V\frac{{\rm d} T_{\rm s}(t)}{{\rm d}t}.
\end{equation}
Hence, we obtain
\begin{align}
        \frac{{\rm d} T_{\rm s}(t)}{{\rm d}t}= -\beta\sigma  K_{\rm o}(K)(T_{\rm s}(t)-T_{\rm a}(d)),
\label{eq:Ts1}
\end{align}
where $\beta=\rho_{\rm a}c_{\rm a}/\rho_{\rm s}c_{\rm s}V$. 
Therefore, for the initial skin temperature, $T_{\rm s0}$, the differential equation (\ref{eq:Ts1}) can be solved and $T_{\rm s}(t)$ can be calculated as
\begin{eqnarray}\label{eq:coolingModel}
    T_{\rm s}(t)=T_{\rm s0}+\left(T_{\rm a}(d)-T_{\rm s0}\right)\left(1-e^{-\beta\sigma  K_{\rm o}(K)t}\right).
\end{eqnarray}
The change in skin temperature is related to the output volume flow rate, $K$, of cold air and the distance, $d$, between the skin and the cold air outlet. By using (\ref{eq:coolingModel}), the skin temperature change can be estimated to elicit target temperature changes by adjusting the volume flow rate of the cold air. 

\section{System and Model Evaluation}
\par We measured the temperature changes according to the volume flow rate of cold air at different distances to evaluate the prototype and proposed model for eliciting cold sensations.
\subsection{Experiment 1—Evaluation Using Phantom}
\par Owing to the complex structure of the skin and its variety of thermal properties, even under the same conditions, measurement results may vary. Furthermore, conducting multiple measurements on the same area may harm or cause discomfort to a participant. Therefore, we first measured the temperature change using a silicone sheet as a phantom.
\subsubsection{Experimental Conditions and Settings}
\par
\begin{figure}[t]
	\centering
	\includegraphics[scale=0.27]{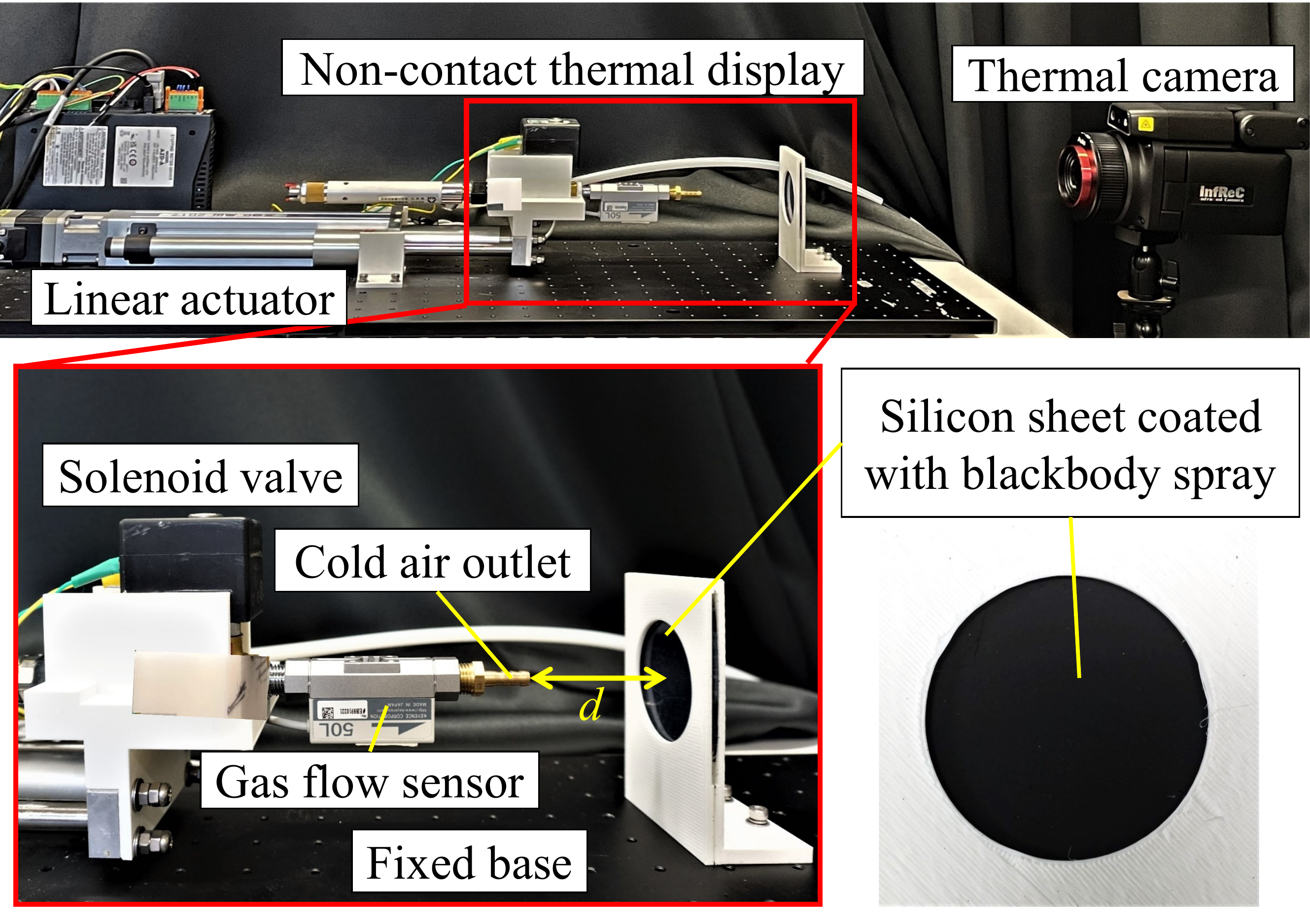}
	\caption{Experimental settings for evaluation using phantom.}
	\label{fig:siliconeExperimentsSetting}
\end{figure}

 Fig.~\ref{fig:siliconeExperimentsSetting} shows the experimental settings for the model evaluation using the phantom. The silicone sheet was positioned directly opposite the cold air outlet. To vary the distance between the cold air outlet and the phantom, we mounted the non-contact cooling prototype on a linear actuator (Oriental Motor EAC4W-D15-AZAA). To observe whether the cold air output meets the target volume flow rate, we installed a gas flow sensor (Keyence FD-A50) between the solenoid valve and the cold air outlet. To measure the temperature distribution on the silicone sheet, we placed a thermal camera (Avionics InfReC R450 with temperature resolution $\leq$0.025~$^\circ$C) on the opposite side of the silicone sheet. We applied blackbody paint (TASCO TA410KS with emissivity of 0.94) to one side of the silicone sheet facing the thermal camera to obtain an accurate temperature distribution measurement. The thickness of the silicone sheet was 1~mm. We measured the temperature distribution of a circular area with a diameter of 50~mm on the surface of the silicone sheet. As the sheet thickness was only 1~mm, the heat of the surface in contact with the cold air was quickly transferred to the other side, resulting in a negligible temperature error on both sides.
\par We evaluated two sets of conditions. First, we fixed the distance to 35~mm and set the volume flow rates of the generated cold air to 8, 16, 24, 32, and 40~L/min. Then, we fixed the volume flow rate to 32~L/min and set the distances to 5, 35, 65, 95, and 125~mm. We performed 10 measurements per condition. Prior to each measurement, we allowed the silicone sheet to rest for 90~s to stabilize its initial temperature. Then we delivered a cold stimulus for 6~s while measuring the temperature distribution. The environmental temperature was adjusted to 24~$^\circ$C using an air conditioner, and the generated cold air temperature at the cold air outlet was 0~$^\circ$C, as indicated by a temperature sensor (Sensirion SHT85). We considered the silicone sheet to be a cylinder with a radius of 25~mm and a height of 1~mm, thus obtaining a volume of 1963.5~mm$^3$. The characteristics of the silicone sheet and air at 0~$^\circ$C are listed in Table~\ref{table:theCharacteristics}~\cite{Chemical}.
\begin{table}[t]
	\caption{Characteristics of air at 0~$^\circ$C, silicone, and skin~\cite{Chemical,skin}}
	\label{table:theCharacteristics}
	\begin{center}
		\begin{tabular}{|c|c|c|} \hline
			Substance & Density & Specific heat capacity \\ \hline \hline
			Air at 0~$^\circ$C& 1.32~kg/m$^{3}$&1007~kJ/kg$\cdot$℃\\\hline
			Silicone & 970~kg/m$^{3}$&1600~kJ/kg$\cdot$℃\\ \hline
			Skin & 1200~kg/m$^{3}$ & 3600~kJ/kg$\cdot$℃\\ \hline
		\end{tabular}
	\end{center}
\end{table}

\subsubsection{Results}
\par
\begin{figure*}[t]
	\centering
	\subfloat[]{\includegraphics[scale=0.24]{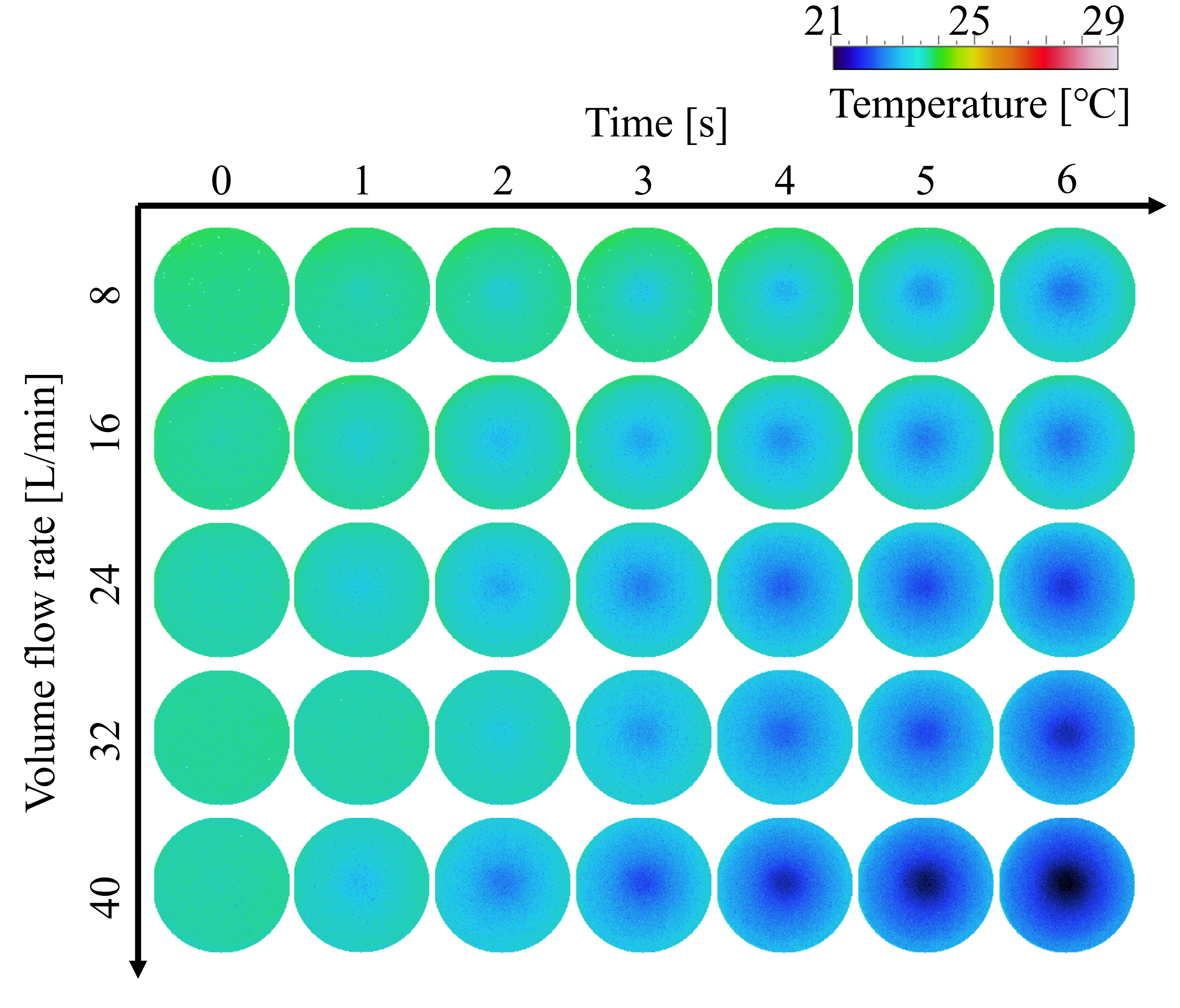}}
	\hfil
	\subfloat[]{\includegraphics[scale=0.24]{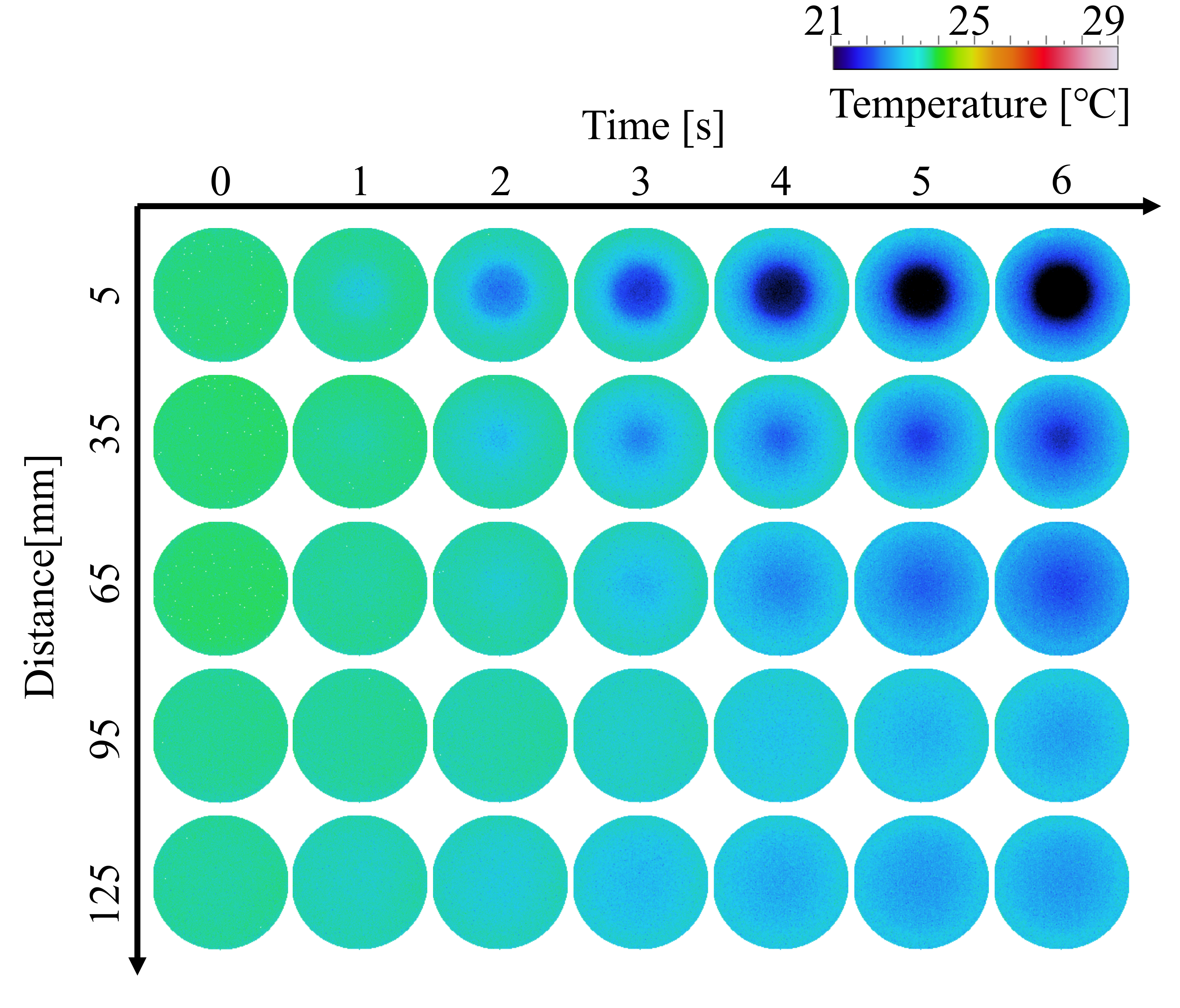}}
	\hfil
	\subfloat[]{\includegraphics[scale=0.7]{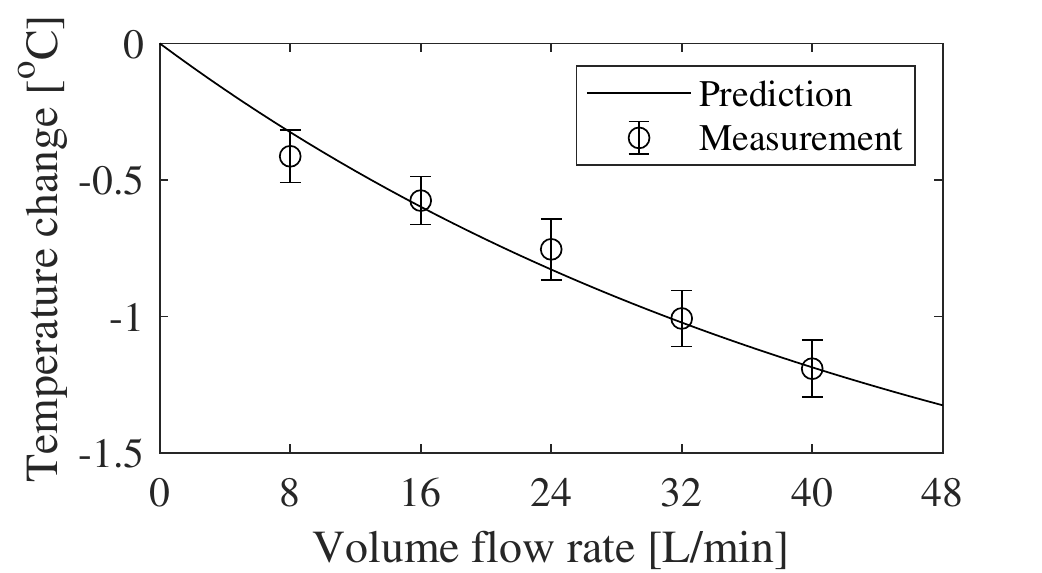}}
	\hfil
	\subfloat[]{\includegraphics[scale=0.7]{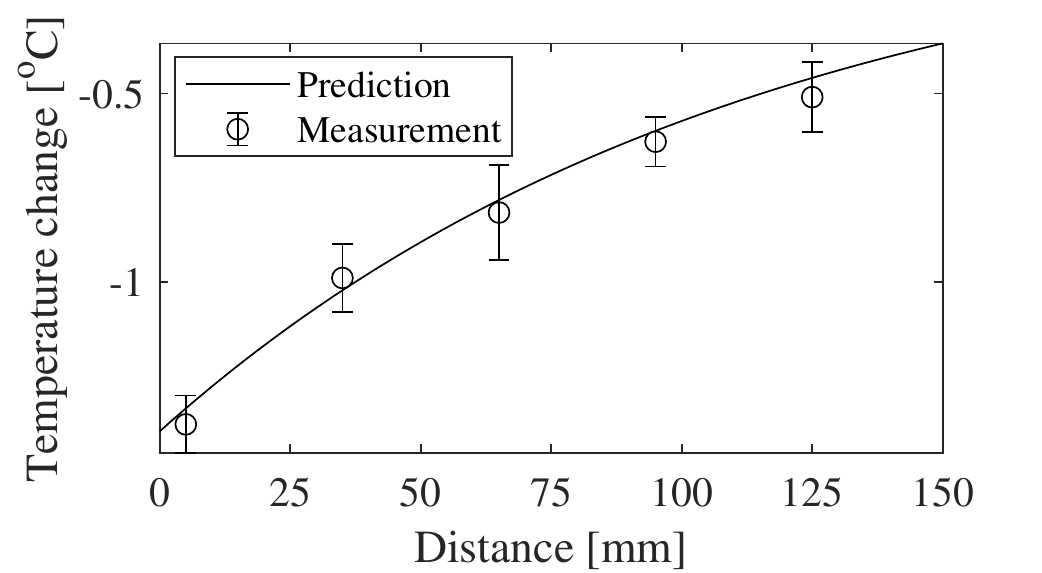}}
	\caption{Experimental results using silicone phantom. Temperature distributions over 6~s at different volume flow rates at a fixed distance (a) and different distances at a fixed volume flow rate (b). Temperature changes over 6~s at different volume flow rates at a fixed distance (c) and different distances at a fixed volume flow rate (d).}
	\label{fig:resultUsingPhantom}
\end{figure*}

 Figs.~\ref{fig:resultUsingPhantom}(a) and (b) show temperature distributions of the silicone sheet measured at a fixed distance and a fixed volume flow rate, respectively. Each circle represents the temperature distribution of the measured circular area on the silicone sheet surface at the corresponding parameter over time. The results show that the temperature change increases as the volume flow rate increases or as the distance decreases. In our cooling model, coefficients $K_{\rm m}$, $\alpha$, $\theta$, and $\sigma$ are unknown. Using the measurement results for both sets of conditions, we estimated the coefficients using the iterative least-squares method, obtaining $\sigma \cdot K_{\rm m}=4.91\times10^{-5}~{\rm m^3/s}$, $\alpha=8.91~{\rm m^{-1}}$, and $\theta=1.18\times10^{3}~{\rm s/m^3}$ with a root mean-square error (RMSE) of 0.08~$^\circ$C for all measurements. Figs.~\ref{fig:resultUsingPhantom}(c) and (d) show the predicted results and measurement average of the temperature changes over 6~s at a fixed distance and a fixed volume flow rate, respectively. 
\subsection{Experiment 2—Evaluation on Skin}
\par \textcolor{black}{We also verified whether the proposed model could accurately estimate changes in the skin temperature.}
\subsubsection{Participants}
\par For this experiment, 15 paid participants (aged 20--26 years, 2 females and 13 males) were enrolled. Each participant received approximately USD 9 (¥1,000) in the form of an Amazon gift card as monetary compensation. The recruitment of participants and experimental procedures were approved by the Ethical Committee of the Faculty of Engineering, Information, and Systems, University of Tsukuba, Japan (approval number 2020R383). All participants provided written informed consent prior to taking part in the study.
\subsubsection{Experimental Conditions and Settings}
\par
\begin{figure}[t]
	\centering
	\includegraphics[scale=0.28]{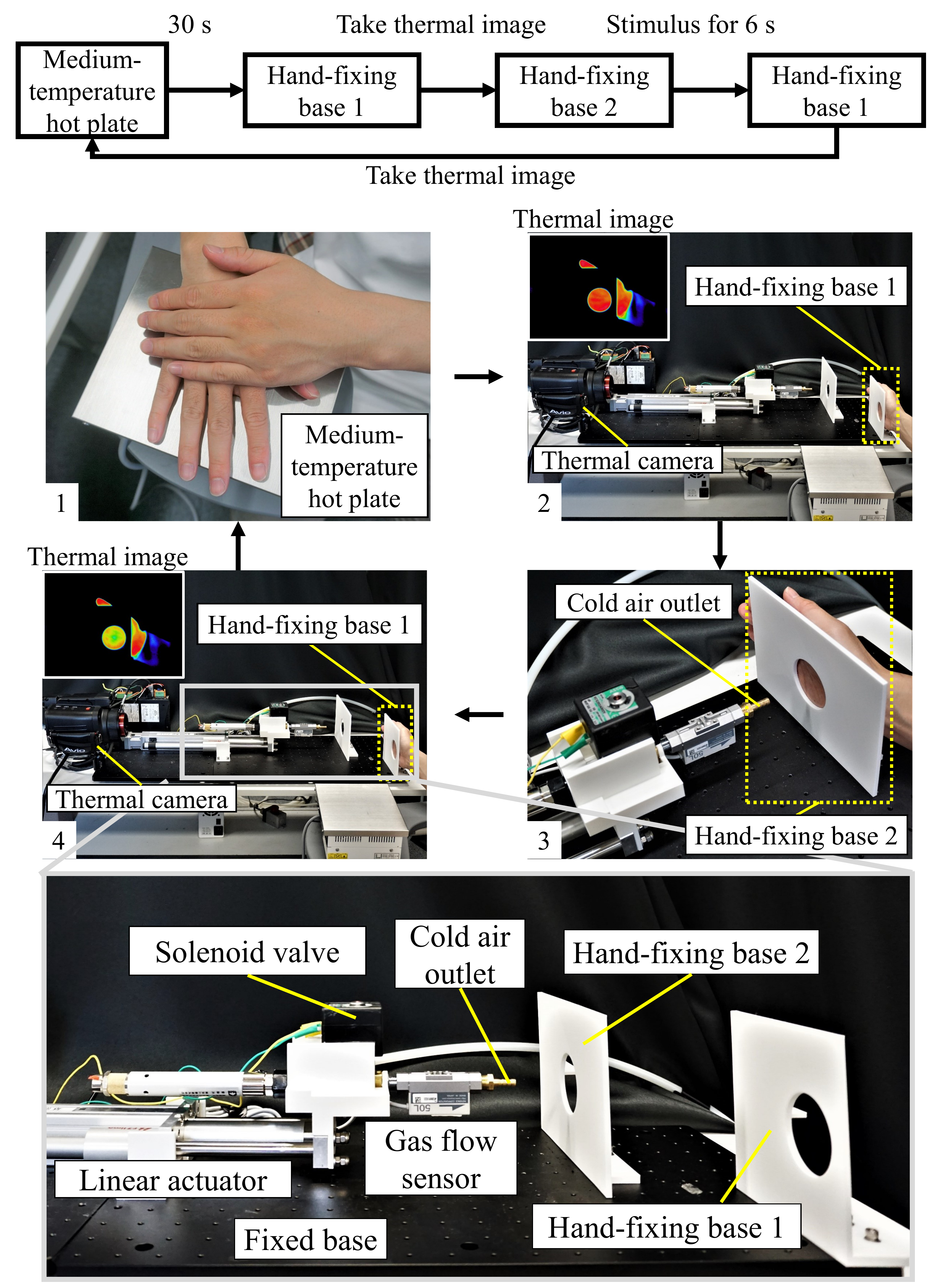}
	\caption{Experimental settings for evaluation on skin.}
	\label{fig:SkinExperimentsProcedure}
\end{figure}

 As shown in Fig.~\ref{fig:SkinExperimentsProcedure}, \textcolor{black}{we chose to provide stimuli perpendicularly to the palm of the right hand (i.e., glabrous, non-hairy skin).} We set up two hand-fixing bases with holes (circles with a radius of 25~mm) to regulate the size of the skin area. We placed base 1 facing the thermal camera. When the participants placed their hands on it, we measured the temperature distribution of their hands using a thermal camera. We placed base 2 facing the cold air outlet. When the participants placed their hands on it, they received a cold stimulus. A medium-temperature hot plate was placed on the participants’ chair. We asked them to place their hands on it before each measurement to ensure a similar initial temperature. Prior to the experiment, we asked the participants to sit still for five minutes to measure the resting baseline temperature of their skin and set the medium-temperature hot plate at the resting baseline temperature.
\par
\begin{figure}[t]
	\centering
	\includegraphics[scale=0.2]{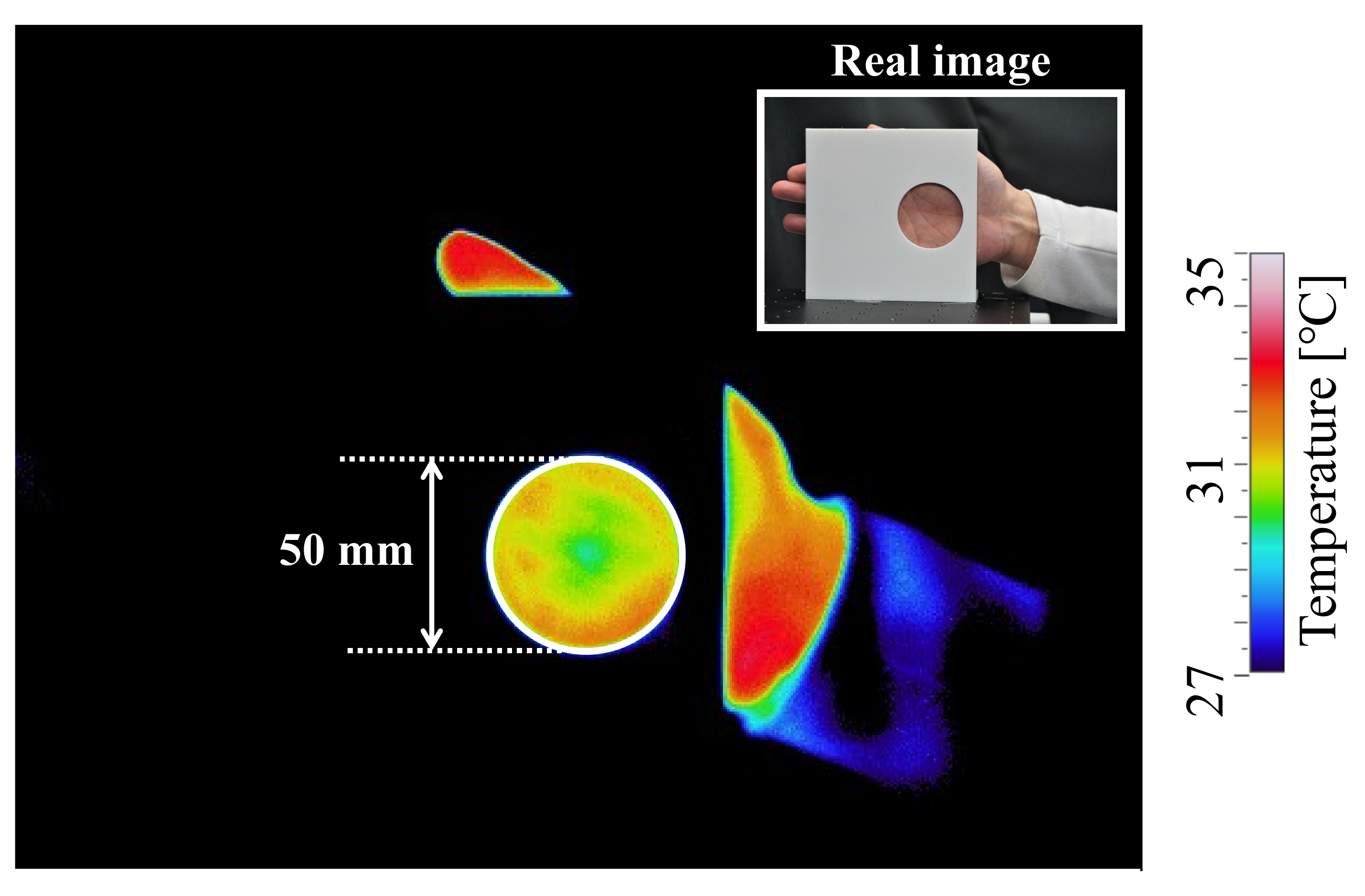}
	\caption{Measurement area of skin.}
	\label{fig:measurementArea}
\end{figure}

 We conducted experiments under the two sets of conditions used for evaluation using the phantom (experiment 1). For each participant, we measured the temperature changes created by the same stimulus three times. In each measurement, we asked participants to place their hands on the hot plate for 30~s. Next, we asked them to place a hand on hand-fixing base 1 to capture a thermal image. Then, we asked them to place a hand on hand-fixing base 2 to present the cold stimulus for 6~s. Finally, we asked them to place their a back on hand-fixing base 1 to capture another thermal image. We adjusted the environmental temperature to 24~$^\circ$C using an air conditioner, while the temperature of the generated cold air at the outlet remained at 0~$^\circ$C, as indicated by a temperature sensor. As shown in Fig.~\ref{fig:measurementArea}, we calculated the average temperature inside the white circle (approximate diameter of 50~mm) from the thermal image. As cold receptors are located directly under the epidermis at a depth of 0.15 to 0.17~mm~\cite{coldRecepter}, we assumed the thickness of the skin tissue in contact with cold air to be 0.2~mm. Therefore, we considered the skin tissue to be a cylinder with a radius of 25~mm and a height of 0.2~mm, obtaining a volume of 392.7~mm$^3$. The characteristics of the skin and air at 0~$^\circ$C are listed in Table~\ref{table:theCharacteristics}~\cite{skin}.
\subsubsection{Results}
\par
\begin{figure*}[t]
	\centering
	\subfloat[]{\includegraphics[scale=0.24]{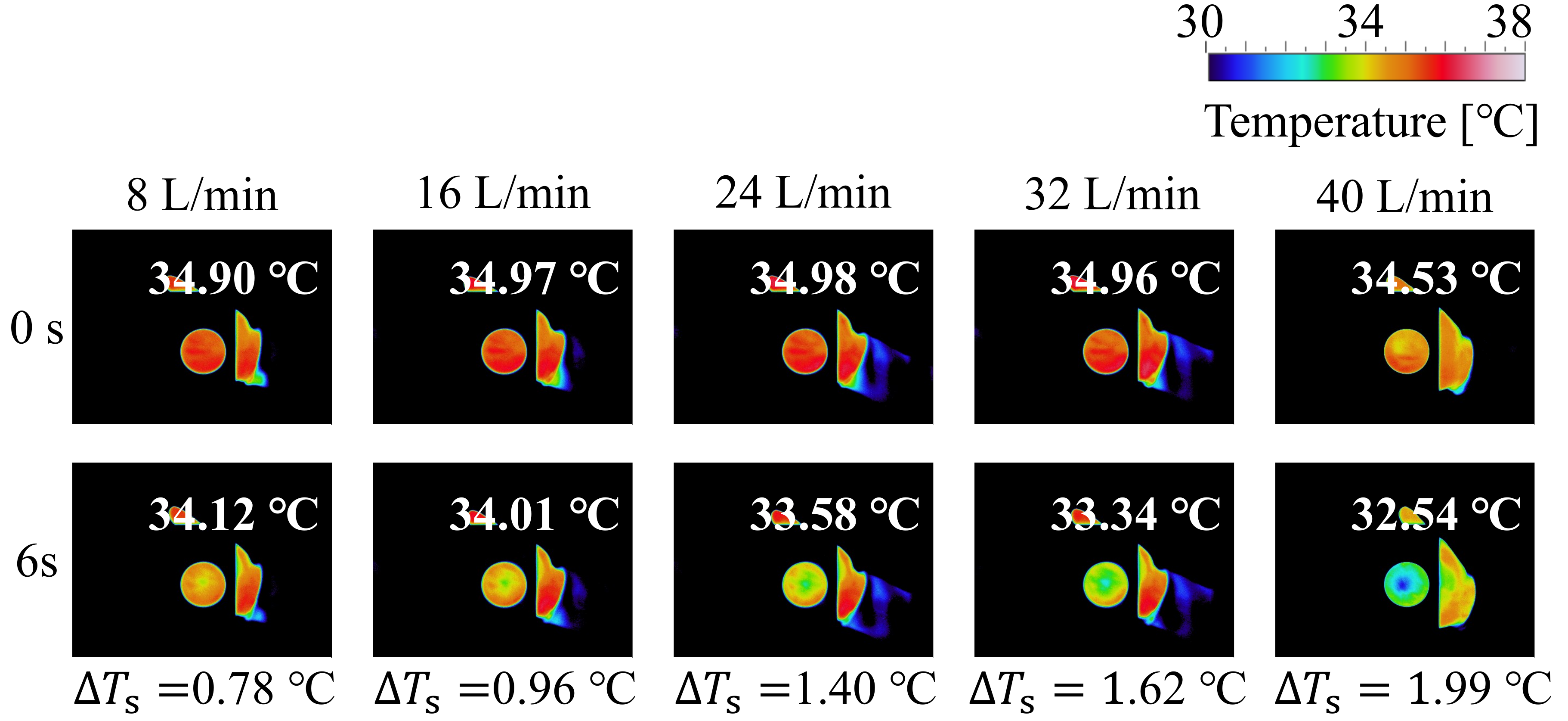}}
	\hfil
	\subfloat[]{\includegraphics[scale=0.24]{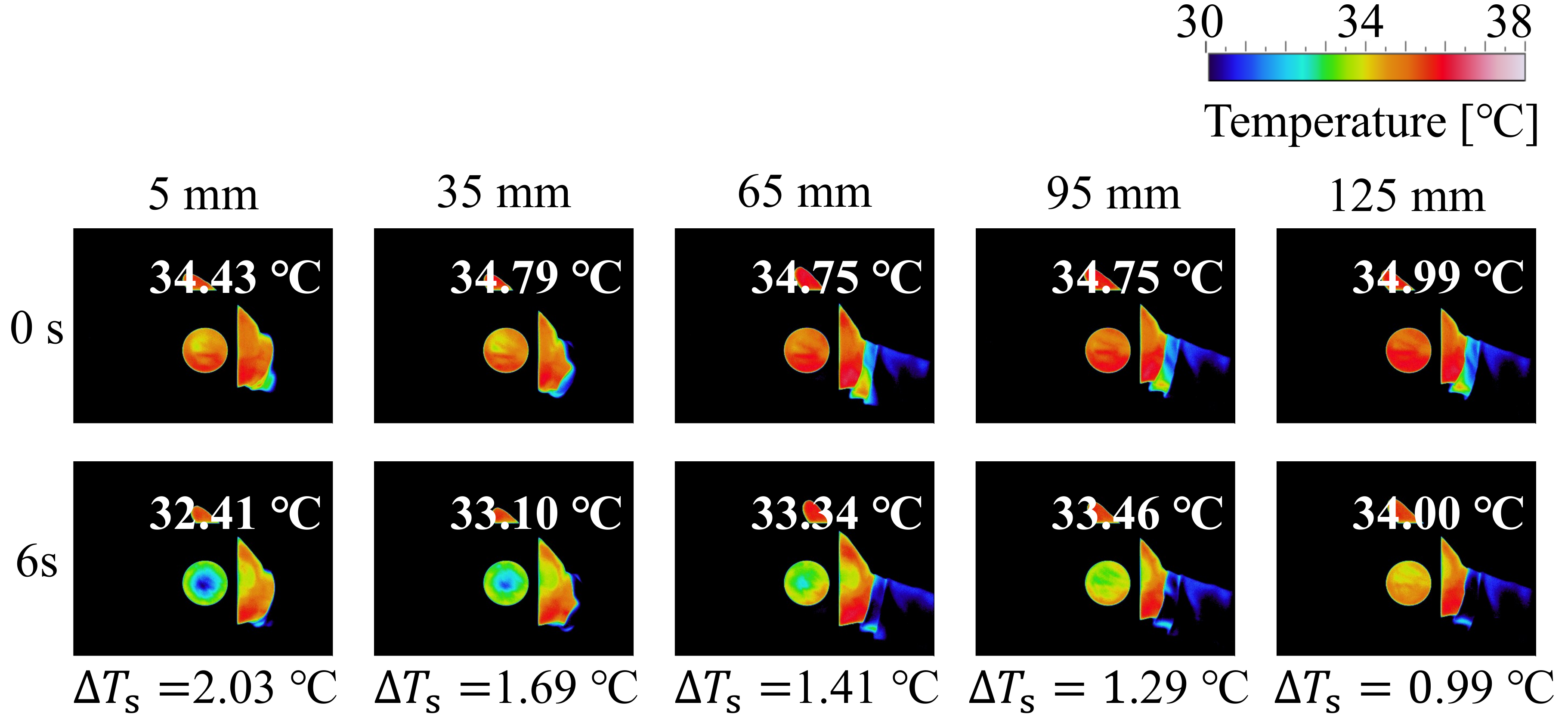}}
	\hfil
	\subfloat[]{\includegraphics[scale=0.7]{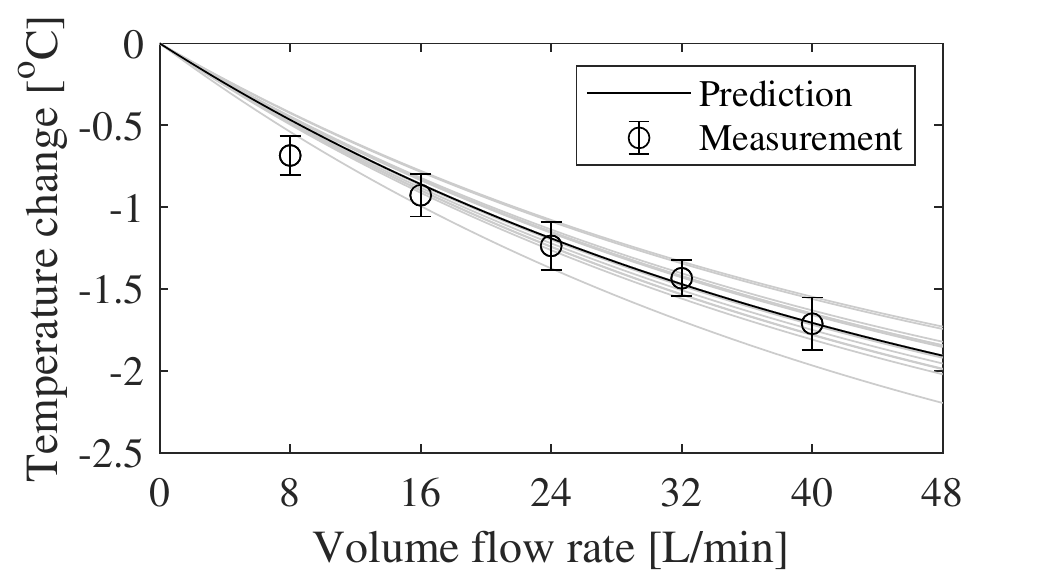}}
	\hfil
	\subfloat[]{\includegraphics[scale=0.7]{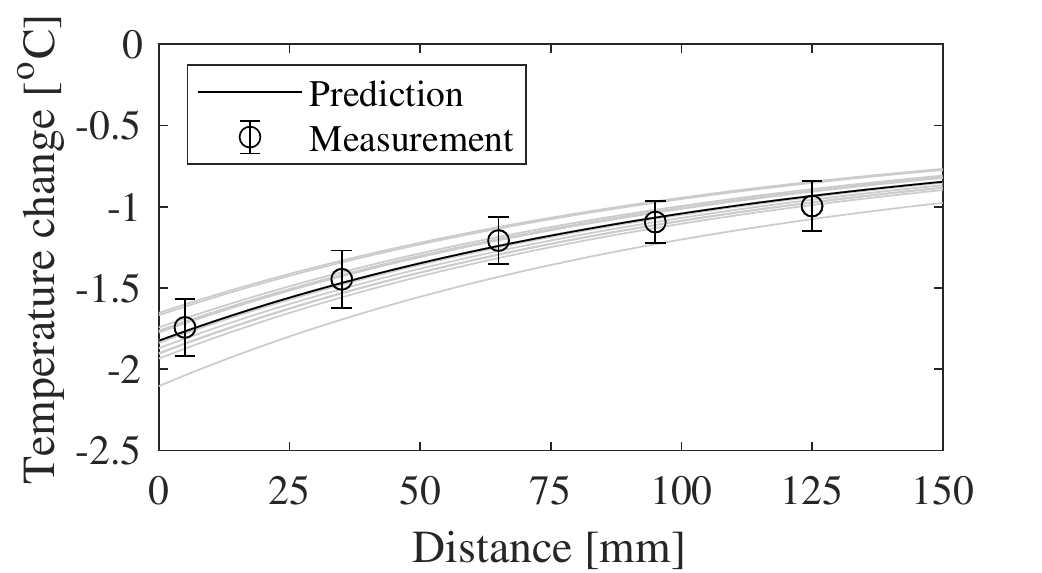}}
	\caption{Experimental results on skin. Temperature distributions over 6~s at different volume flow rates at a fixed distance (a) and different distances at a fixed volume flow rate (b). Temperature changes over 6~s at different volume rates at a fixed distance (c) and different distances at a fixed volume flow rate (d).}
	\label{fig:resultUsingSkin}
\end{figure*}

 Figs.~\ref{fig:resultUsingSkin}(a) and (b) show the experimental results at fixed distances, in which the skin temperature change increases as the volume flow rate increases, and at a fixed volume flow rate, in which the skin temperature change decreases as the distance increases, respectively. The absorption ratio, $\sigma$, of the skin may differ from that of the silicone sheet. Therefore, we applied the iterative least-squares method to estimate coefficients $\sigma \cdot K_{\rm m}$. Meanwhile, we assumed that the other coefficients were the same as those obtained in experiment 1. We obtained $\sigma \cdot K_{\rm m}=2.59\times10^{-5}~{\rm m^3/s}$ with an RMSE of 0.16~$^\circ$C for all measurements. Figs.~\ref{fig:resultUsingSkin}(c) and (d) show the predicted and measured skin temperature changes over 6~s at a fixed distance and a fixed volume flow rate, respectively. The solid black line represents the predicted average result using our cooling model for all participants, and each solid gray line represents a prediction using our cooling model per participant. The circles and error bars represent the measurement results. The experimental results show that our cooling model can suitably fit temperature changes for volume flow rates above 16~L/min at distances below 125~mm. 

\section{Psychophysical Evaluation}
\par To investigate the influence of the proposed system on human perception, we conducted two experiments and evaluated the performance of the prototype system in psychophysical cold discrimination. Specifically, we examined how the volume flow rate of cold air (experiment 3) and the distance between the cold air outlet and skin (experiment 4) affected cold sensations. Furthermore, we determined thresholds for eliciting cold sensations without contact. 

\subsection{Participants}
\par The 15 participants from experiment 2 also took part in experiments 3 and 4.
\subsection{Experimental Conditions and Settings}
\par
\begin{figure}[t]
	\centering
	\includegraphics[scale=0.28]{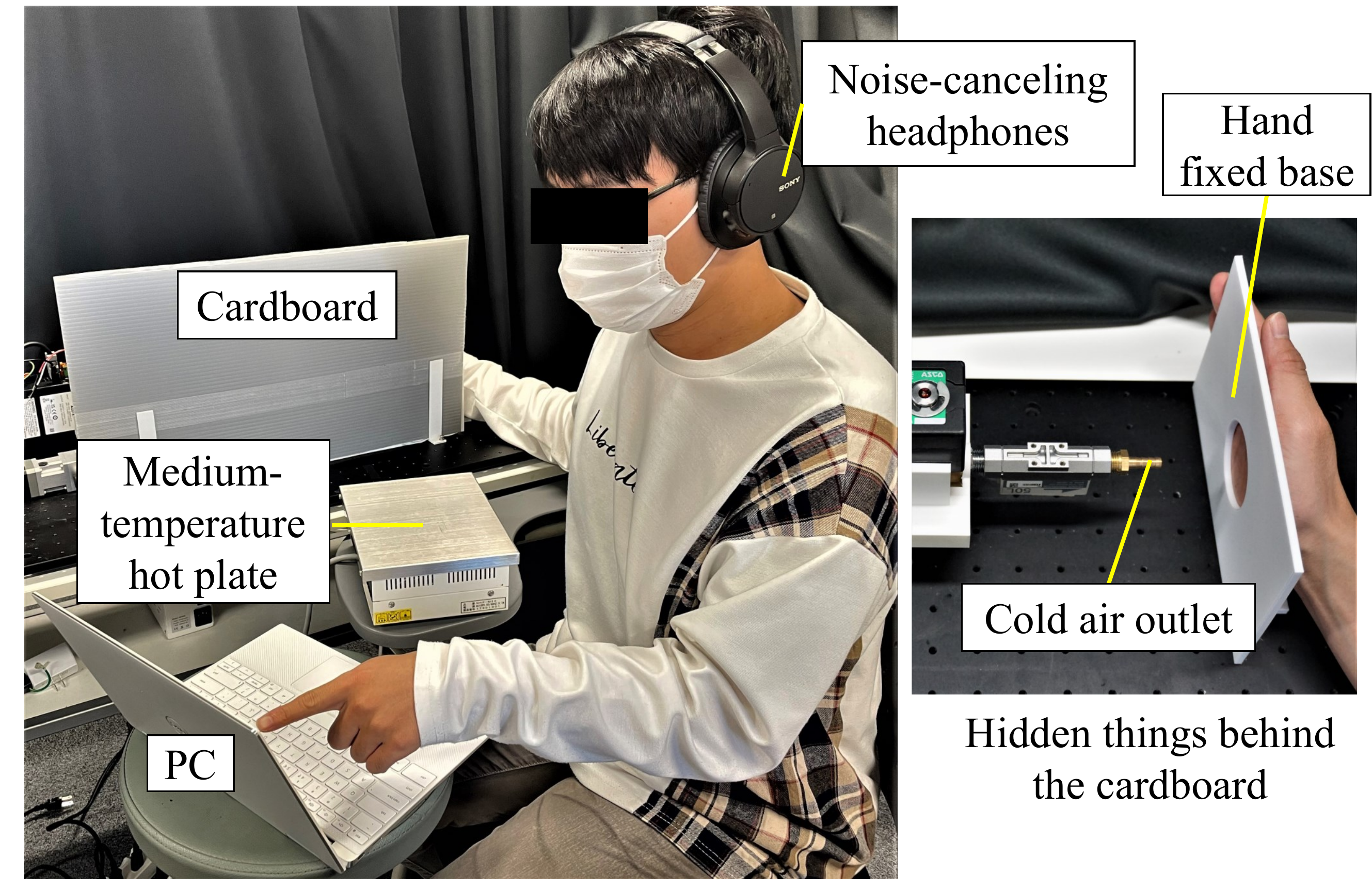}
	\caption{Experimental settings for psychophysical evaluation.}
	\label{fig:PsychologicalEvaluationSetting}
\end{figure}

 Fig.~\ref{fig:PsychologicalEvaluationSetting} shows the experimental settings for experiments 3 and 4. To prevent the participants from observing the distance between the cold air outlet and their hand and to reduce the influence of environmental airflow, we covered the prototype with cardboard. To avoid the influence of sounds from the compressor, solenoid valve, and environment, every participant wore a pair of noise-canceling headphones and listened to white noise. In the experiments, a one-hand fixing base was installed so that the hand would face the cold air outlet as it provided the cold stimuli. \textcolor{black}{We also chose to provide stimuli perpendicularly to the palm of the right hand (i.e., glabrous, non-hairy skin)} and set up hand-fixing bases with a hole (a circle with a radius of 25~mm) to regulate the size of the skin area. The participants were asked to respond to a questionnaire about the perceived cold sensations through keyboard input on a computer. We provided a medium-temperature hot plate at the resting baseline temperature of their skin. We adjusted the environmental temperature to 24~$^\circ$C using an air conditioner, and the temperature of the generated cold air at the outlet was 0~$^\circ$C, as indicated by a temperature sensor.
\subsection{Experiment 3—Influence of Volume Flow Rate on Cold Sensation}
\subsubsection{Experimental Procedure}
\par We used the forced-choice paradigm with constant stimuli and evaluated the discrimination sensitivity to coldness of the participants. To investigate the influence of the volume flow rate on cold sensation, we set the fixed distance (35~mm), the standard stimuli to a volume flow rate (24~L/min), and the comparison stimuli (8, 16, 24, 32, and 40~L/min). Each comparison stimulus was randomly presented for 10 trials, resulting in 50 trials (5 stimuli $\times$ 10 trials). 
In each trial, we presented the first stimulus for approximately 1.5~s; then, the system stopped presenting stimuli for approximately 1~s. We presented the subsequent stimulus for 1.5~s. After presenting these two stimuli, the participants used the keyboard to indicate which stimulus was colder. We randomly determined the order of the standard and comparison stimuli. To avoid unreliable answers due to fatigue, the participants took a break every 5 trials, in which they were allowed to remove their hands from the base and rest them on the medium-temperature hot plate for 30~s to restore the initial temperature. 
\begin{figure*}[t]
	\centering
	\subfloat[]{\includegraphics[scale=0.59]{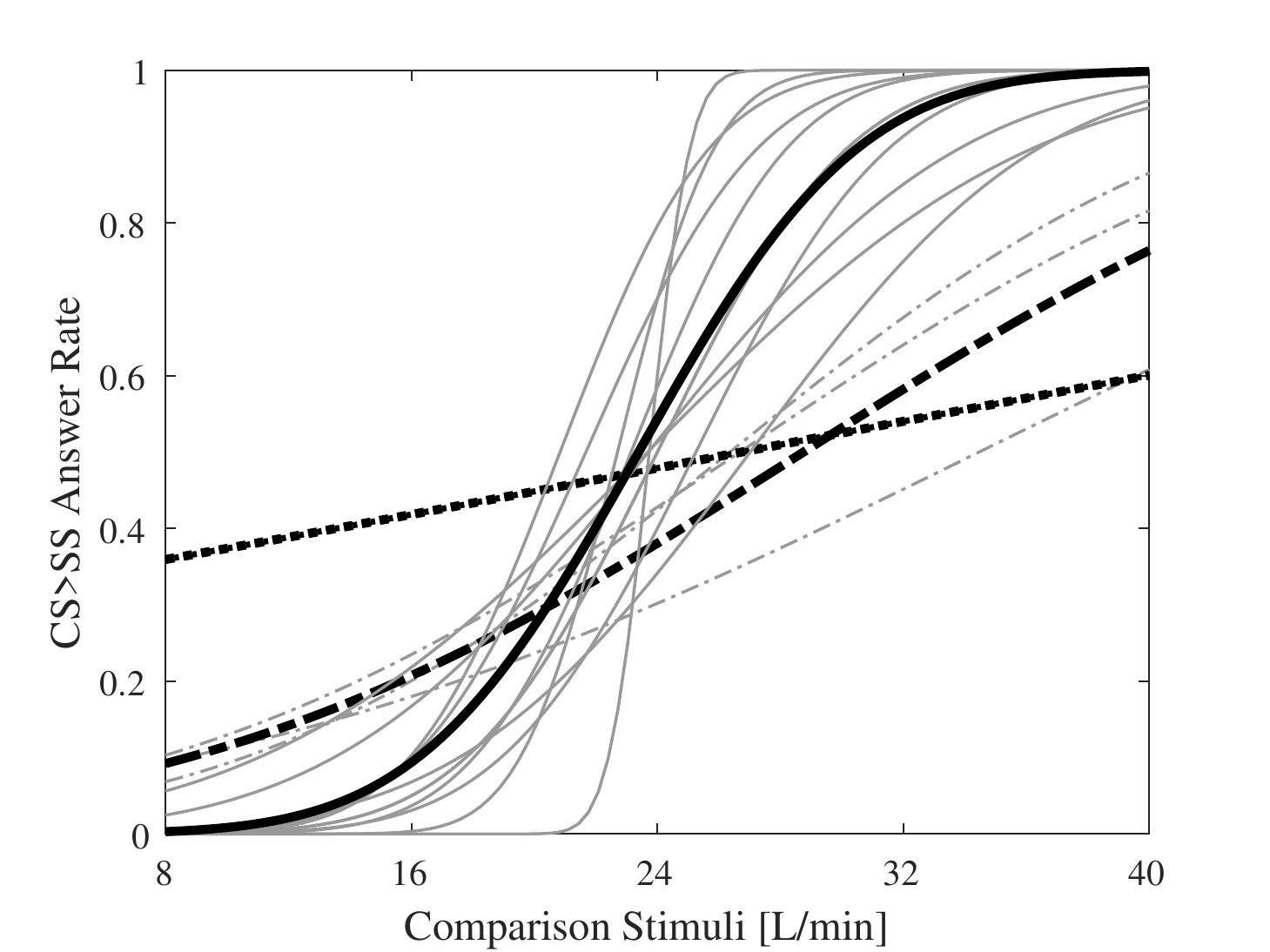}}
	\subfloat[]{\includegraphics[scale=0.59]{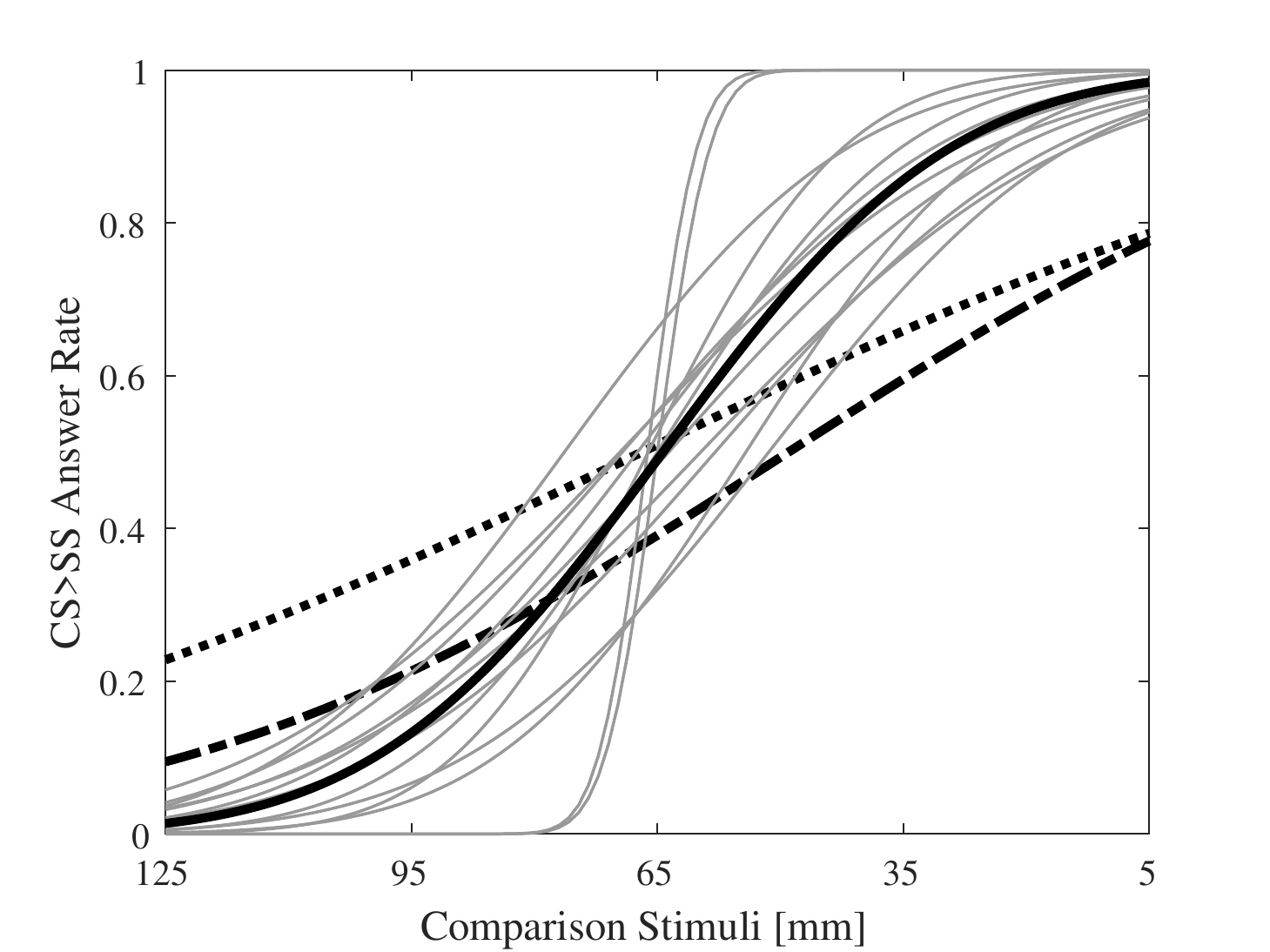}}
	\caption{\textcolor{black}{Probability of comparison being colder from psychophysical evaluation results (CS: comparison stimuli; SS: standard stimuli) for different volume flow rates (a) and different distances (b). Each gray line represents a single subject’s data. The gray solid lines indicate successful discrimination of cold stimuli, gray dash-dotted lines indicate chance level of performance for most stimuli, and gray dotted lines indicate failure to discriminate cold stimuli. Black solid, dash-dotted, and dotted lines represent the average results of successfully discriminating cold stimuli, the average of chance level of performance for most stimuli, and the average of failure to discriminate cold stimuli, respectively.}}
	\label{fig:FlowVolumeRateFixed}
\end{figure*}
\subsubsection{Results}
\par To evaluate the discrimination performance, we calculated the probability of a participant finding the comparison stimulus colder across stimulus pairs. Fig.~\ref{fig:FlowVolumeRateFixed}(a) shows the psychometric curves fit to these probabilities using the normalized cumulative distribution function at different flow volume rates. 

\par To quantitatively evaluate the discrimination performance, we obtained the point of subjective equality (PSE) and the just-noticeable difference (JND) as half the difference between the inverse of the curves at 0.75 and 0.25 thresholds. The mean PSE and JND of the results with s-curve shape (the black solid line in Fig.~\ref{fig:FlowVolumeRateFixed}(a)) were 23.61~$\pm$~1.57~L/min and 6.82~$\pm$~3.23~L/min, respectively. The mean PSE and JND of the overall results were 24.93~$\pm$~3.11~L/min and 13.94~$\pm$~16.44~L/min, respectively.

\subsection{Experiment 4—Influence of Distance on Cold Sensation}
\subsubsection{Experimental Procedure}
To investigate the influence of the distance on cold sensation, we set a fixed volume flow rate (32~L/min) and followed the same procedure as in Experiment 3. We set the standard stimulus to 65~mm and the comparison stimuli to 5, 35, 65, 95, and 125~mm. 
\subsubsection{Results}
\par As with experiment 3, we calculated the probability of a participant finding the comparison stimulus colder across stimulus pairs. Fig.~\ref{fig:FlowVolumeRateFixed}(b) shows the psychometric curves fit to these probabilities using the normalized cumulative distribution function at different distances.
\par We also calculated PSEs and JNDs as in Experiment 3. The mean PSE and JND of the results with s-curve shape (the black solid line in Fig.~\ref{fig:FlowVolumeRateFixed}(b)) were 66.01~$\pm$~6.47~mm and 34.11~$\pm$~13.37~mm, respectively. The mean PSE and JND of all the lines were 66.82~$\pm$~7.15~mm and 41.77~$\pm$~23.68~mm, respectively.

\section{Discussion}
\par Unlike previous studies, we focused on eliciting thermal sensations without contact to create a new way of 
recreating interactions. To this end, we developed a method for eliciting non-contact intensity-adjustable cold sensations based on the vortex effect. Using this effect, we generated low-temperature cold air (0~$^\circ$C) and transferred it directly to the skin of a participant for rapid cooling. The results of experiment 3 \textcolor{black}{demonstrate that the intensity of cold sensations varies as a function of the volume flow rate of the generated cold air}  reaching the skin. \textcolor{black}{Our presentation component, i.e., a cold air outlet (a round tube with a diameter of 8~mm), is lightweight} and can be miniaturized and customized for easy incorporation into VR wearable devices. In addition, the compressor can be placed in a remote location to resolve problems related to size and noise.
\par We developed a cooling model that integrates the characteristics of the proposed method and relates changes in skin temperature to the volume flow rate of cold air and the distance between the cold air outlet and skin. As shown in Fig.~\ref{fig:resultUsingSkin}, although \textcolor{black}{inter-subject variability} exists, the variations were smaller than the temperature range. Consequently, the model can suitably predict approximately the same results with \textcolor{black}{an} RMSE of 0.16~$^\circ$C. According to the measurement results, our prototype can realize a maximum temperature change of 0.34~$^\circ$C/s, which is sufficient to elicit a strong cold sensation~\cite{thermoreceptors}. 
\par Section \uppercase\expandafter{\romannumeral4} presents the estimation, measurement, and evaluation of the cooling model using the average temperature change in an area. The corresponding results show that the cooling model can predict the temperature change of a phantom with an RMSE of 0.08~$^\circ$C. For human skin, the cooling model can predict the trend of temperature changes for a volume flow rate of $16$ to $40$~L/min at $5$ to $125$~mm with an RMSE of 0.16~$^\circ$C. We attribute these errors to individual differences in thermal properties of the skin among other factors. In addition, as the hand temperature is not measured continuously, the time delay of moving the hand may also lead to errors. The temperature change is larger than the predicted results when the volume flow rate is 8~L/min (i.e., the weakest cold stimulus). 
\par \textcolor{black}{When the cold stimulus is strong, an initial response is a strong vasoconstriction, leading to a rapid decrease in temperature. Meanwhile, paradoxical and cyclical vasodilation occurs and enables large amounts of blood to pass and convey heat to the surrounding tissue to compensate for changes in the skin temperature~\cite{coldvasodilation,coldvasodilation1,coldvasodilation2}. }
In contrast, when a cold stimulus is not sufficiently strong, the body does not generate excessive heat in response. In other words, there is a weak resistance to skin temperature changes in this case. Therefore, the temperature changes were more considerable when the cold stimulus was relatively weak. We plan to consider this effect in future to improve the accuracy of our cooling model. Furthermore, as mentioned in Section \uppercase\expandafter{\romannumeral3}, the area of the skin directly in contact with cold air may differ according to the volume flow rate and distance. In future work, we will determine the contact area and analyze its effects.
\par Section \uppercase\expandafter{\romannumeral5} reports the performance of our prototype in psychophysical cold discrimination experiments. The discrimination threshold for volume flow rate was 6.82~L/min (corresponding to a temperature change rate difference of 0.05~$^\circ$C/s), and that for distance was 34.11~mm (corresponding to a temperature change rate difference of 0.04~$^\circ$C/s). Moreover, \textcolor{black}{participants who had difficulty separating different stimuli (the dash-dotted and dotted lines in Fig.~\ref{fig:FlowVolumeRateFixed})} had a higher resting baseline temperature of approximately 35~$^\circ$C. They may require a larger temperature change to feel cold. In future, we plan to increase the maximum temperature change allowed by our device to further study human thermal perception to improve the reliability of the proposed method for general applications involving thermal stimulation. 
\par Changes in the volume flow rate and distance cause different pressure sensations on the skin. The intensity of such sensations may affect the participants' ability to distinguish cold sensations. In future work, we will use nets to weaken pressure sensations by turning cold airflow into a turbulence. Then, we will conduct comparative experiments to examine the influence of pressure sensations on temperature perception. \textcolor{black}{On the other hand,  the synthetic sensation of wetness is thought to be produced from a combination of specific skin thermal and tactile inputs, registered through thermoreceptors and mechanoreceptors, respectively\cite{wetness}. Therefore, it is also possible to generate a sensation of wetness by varying pressure sensations and thermal sensations on the skin. In future, we plan to investigate this possibility as well.}
\par In this study, we conducted experiments on the skin of the hands. We plan to test this stimulus on other sensitive areas of the body, such as the cheek, to investigate cold sensation perception and determine the smallest area of the skin that should be cooled to elicit a cold sensation for users to experience a cold environment realistically. Furthermore, we will incorporate the presentation of hot sensations to expand the scope of application. For instance, using electromagnetic waves to simulate the heat of the sun, or combining with hot air to simulate all types of environments from arctic regions to deserts. 

\section{Conclusion}
\par We proposed a method to elicit intensity-adjustable non-contact cold sensations on the human skin using the vortex effect, in which convection is the dominant factor affecting the temperature. To adjust the intensity of the cold sensation, we varied the volume flow rate of the cold air generated using the vortex effect. We also introduced a cooling model that integrates the characteristics of the proposed method, relating the changes in skin temperature to factors of the volume flow rate of air and the distance from the cold air outlet to the skin. The model can estimate skin temperature changes with \textcolor{black}{an} RMSE of 0.16~$^\circ$C. According to the measurement results, our prototype can realize a maximum temperature change of 0.34~$^\circ$C/s, which is sufficient to elicit strong cold sensations. We evaluated the performance of the prototype using psychophysical cold discrimination experiments. The experimental results show that the prototype can produce cold sensations of different intensities for most participants. We also analyzed the thresholds needed to elicit cold sensations without contact. The discrimination threshold for volume flow rate was 6.82~L/min, and that for distance was 34.11~mm. Therefore, in practice, cold sensations of various intensities can be generated by changing any of these parameters or both simultaneously.
We plan to \textcolor{black}{incorporate} our novel method into a multisensory interactive system to provide users with more immersive experiences in applications, such as those of VR and metaverse.

\begin{IEEEbiography}[{\includegraphics[width=1in,height=1.25in,clip,keepaspectratio]{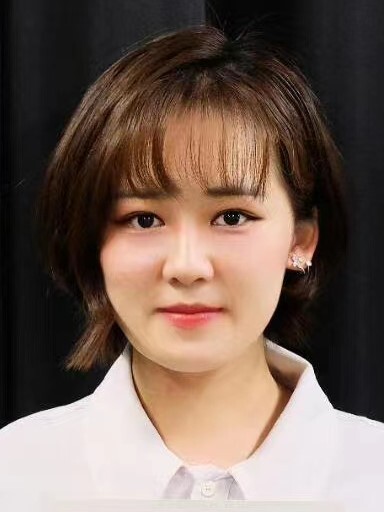}}]{Jiayi Xu} (Student Member, IEEE) received the M.S. degree from the Graduate School of Engineering Science at Osaka University, Osaka, Japan, in 2020. Since 2020, she has been a Ph.D. student with the Graduate School of Science and Technology, Degree Programs in Systems and Information Engineering, University of Tsukuba. Her research interests include haptics with a focus on thermal sensations.
\end{IEEEbiography}

\begin{IEEEbiography}[{\includegraphics[width=1in,height=1.25in,clip,keepaspectratio]{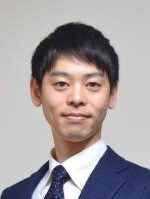}}]{Shunsuke Yoshimoto} (Member, IEEE) received the Ph.D. degree in engineering from Osaka University, Toyonaka, Japan, in 2012. He was an Assistant Professor with the Graduate School of Engineering Science, Osaka University from 2012 to 2019, and a Lecturer with the School of Engineering at The University of Tokyo from 2019 to 2020. Since 2020, he has been a Lecturer with the Graduate School of Frontier Sciences at The University of Tokyo. His research interests include haptic engineering and biomedical instrumentation.
\end{IEEEbiography}

\begin{IEEEbiography}[{\includegraphics[width=1in,height=1.25in,clip,keepaspectratio]{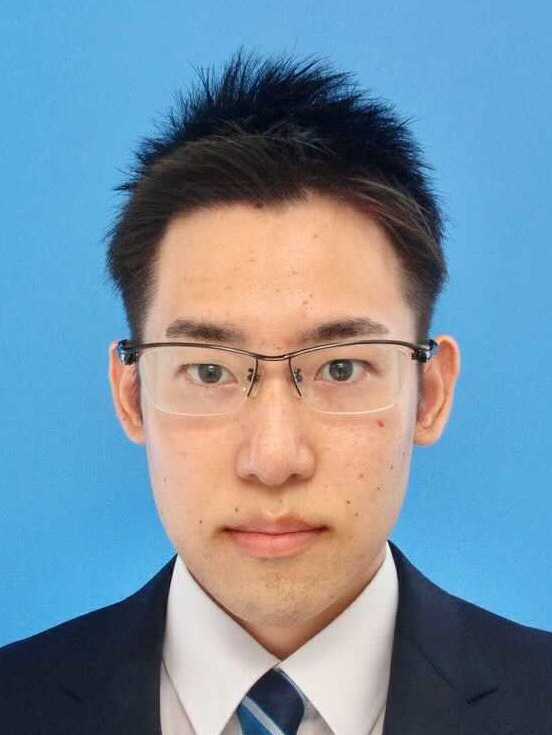}}]{Naoto Ienaga} (Member, IEEE) received the Ph.D. degree in engineering from Keio University, Yokohama, Japan, in 2020. Since 2021, he has been an Assistant Professor with the Faculty of Engineering, Information and Systems at the University of Tsukuba. He is currently working on applications of machine learning and computer vision to solve practical problems, especially in fisheries and occupational therapy.
\end{IEEEbiography}

\begin{IEEEbiography}[{\includegraphics[width=1in,height=1.25in,clip,keepaspectratio]{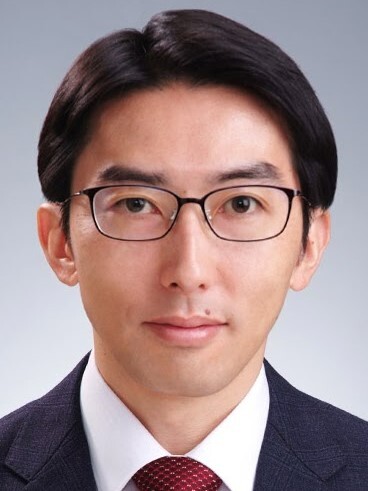}}]{Yoshihiro Kuroda}
(Member, IEEE) received the Ph.D. degree in informatics from Kyoto University, Kyoto, Japan, in 2005. He was an Assistant Professor with the Graduate School of Engineering Science at Osaka University from 2006 to 2013, an Associate Professor with the Cybermedia Center at Osaka University from 2013 to 2016, and an Associate Professor with the Graduate School of Engineering Science at Osaka University from 2016 to 2019. Since 2019, he has been a Professor with the Faculty of Engineering, Information and Systems at the University of Tsukuba. His research interests include haptic interaction technologies and biomedical engineering.
\end{IEEEbiography}
\end{document}